\documentclass[11pt,a4paper]{article}

\usepackage{jcappub}

\usepackage{graphicx}
\usepackage{enumitem}
\usepackage{amsmath}
\usepackage{booktabs}
\usepackage{array}
\usepackage{color}
\usepackage{amssymb}
\usepackage{mathrsfs}
\usepackage[T1]{fontenc}
\usepackage[latin1]{inputenc}
\usepackage[table]{xcolor}  
\usepackage{verbatim}
\usepackage{caption}
\usepackage{subcaption}
\usepackage{rotating}
\usepackage{dcolumn}
\usepackage{bm}

\renewcommand{\d}{\mathrm{d}}
\newcommand{\grad}{\nabla}
\newcommand{\vect}[1]{\bm{\mathrm{{#1}}}}
\newcommand{\im}{\mathrm{i}}

\newcommand{\ipleft}{\langle\kern-0.2em\langle}
\newcommand{\ipright}{\rangle\kern-0.2em\rangle}

\newcommand{\fNL}{f_{\mathrm{NL}}}
\newcommand{\fNLlocal}{\fNL^\text{local}}
\newcommand{\fNLequi}{\fNL^\text{equi}}
\newcommand{\fNLortho}{\fNL^\text{ortho}}
\newcommand{\fNLDBI}{\fNL^{\text{\textsc{dbi}}}}
\newcommand{\fNLEFTone}{\fNL^{\text{\textsc{eft1}}}}
\newcommand{\fNLEFTtwo}{\fNL^{\text{\textsc{eft2}}}}
\newcommand{\fNLghost}{\fNL^{\text{ghost}}}
\newcommand{\lambdaghost}{\lambda^{\text{ghost}}}

\DeclareMathOperator{\Lik}{\mathscr{L}}

\newcommand{\Bref}{B_{\text{ref}}}
\newcommand{\Pzeta}{\mathcal{P}_\zeta}

\newcommand{\Seft}{S_{\text{EFT}}}
\newcommand{\Op}{\mathcal{O}}

\newcommand{\Rmode}{\mathcal{R}}
\newcommand{\Cov}{\hat{\mathcal{C}}}
\newcommand{\Fisher}{\hat{\mathcal{F}}}

\newcommand{\AIC}{\mathrm{AIC}}

\newcommand{\estfNL}{\hat{f}_{\mathrm{NL}}}
\newcommand{\estfNLlocal}{\estfNL^\text{local}}
\newcommand{\estfNLequi}{\estfNL^\text{equi}}
\newcommand{\estfNLortho}{\estfNL^\text{ortho}}

\newcommand{\Mp}{M_{\mathrm{P}}}
\newcommand{\cs}{c_{\mathrm{s}}}
\newcommand{\tildecs}{\tilde{c}_{\mathrm{s}}}
\newcommand{\As}{\mathcal{A}_{\mathrm{s}}}
\newcommand{\cthree}{\tilde{c}_3}

\newcommand{\chisqmle}{\chi^2_{\text{\textsc{mle}}}}

\newcommand{\Nparams}{N_{\lambda}}

\newcommand{\Edecouple}{E_{\text{mix}}}

\renewcommand{\leq}{\leqslant}
\renewcommand{\geq}{\geqslant}
\newcommand{\transpose}{\mathrm{t}}

\DeclareMathOperator{\Or}{O}

\newcommand{\para}[1]{\par\vspace{2mm}\noindent\textbf{{#1}.---}}

\newcolumntype{s}{>{$\displaystyle}l<{$}}
\newcolumntype{t}{>{$\displaystyle}c<{$}}
\newcolumntype{u}{>{$\displaystyle}r<{$}}
\newcolumntype{v}{>{$\displaystyle}m{4cm}<{$}}

\newcolumntype{d}{D{!}{\;\pm\;}{-1}}

\title{Optimal bispectrum constraints on single-field models of inflation}
\author{Gemma J. Anderson,}
\author{Donough Regan}
\author{and David Seery}
\affiliation{Astronomy Centre, University of Sussex,\\
Falmer, Brighton, BN1 9QH, UK}
\emailAdd{G.Anderson@sussex.ac.uk}
\emailAdd{D.Regan@sussex.ac.uk}
\emailAdd{D.Seery@sussex.ac.uk}
\abstract{We use WMAP 9-year bispectrum data to constrain the
free parameters of an `effective field theory'
describing fluctuations in single-field inflation.
The Lagrangian of the theory contains a finite number of
operators
associated with unknown mass scales.
Each operator produces a fixed bispectrum shape,
which we decompose into partial waves
in order to construct a likelihood function.
Based on this likelihood
we are able to constrain four linearly independent
combinations of the mass scales.
As an example of our framework
we specialize our results to the case of
`Dirac--Born--Infeld'
and `ghost' inflation and obtain
the posterior probability for each model,
which
in Bayesian schemes is a useful tool
for model comparison. Our results suggest
that DBI-like
models with two or more free parameters are
disfavoured by the data
by comparison with single-parameter models in the
same class.}

\begin{document}	
\maketitle

\section{Introduction}
Successive microwave-background surveys have accumulated some evidence
for the inflationary paradigm, in which structure in the universe
was seeded by quantum fluctuations during an epoch preceding the hot, dense
phase where nucleosynthesis occurred~\cite{Ade:2013ydc,Hinshaw:2012aka}.
But despite broad support for the overall framework, attempts to identify
the precise degrees of freedom whose quantum fluctuations
were relevant
have met with less success. Whatever microphysics underlay
the putative inflationary epoch remains mysterious.

In scattering experiments, an abundance of observables---including,
among others,
branching ratios, decay rates, and differential dependence on energy or angles---%
allow indirect access to microphysical information
through reconstruction of the correlation functions,
or `$n$-point functions'.
These measure interference between quantum fluctuations
and encode information about the dynamics of the theory.
It is the rich information which can be obtained from
reconstruction of the correlation functions which makes
measurements in particle physics so constraining.

In cosmology our observables are more limited
and
so is the degree to which the $n$-point functions can be reconstructed.
Over a narrow range of scales,
the $n$-point functions of the cosmic microwave background
(`CMB')
anisotropies
are sensitive to
the $n$-point functions of the primordial `curvature perturbation',
which is a calculable,
model-dependent mix of the fluctuations imprinted on
the light fields of the inflationary epoch.
This correspondence has been used for many years
to place
restrictions on the inflationary model space
from measurements of the CMB temperature and
polarization two-point functions.
But if a \emph{three}-point function of the CMB anisotropies could be measured
it would provide access to more nuanced and discriminating microphysical
information.
Ideally we would like to observe systematic
relationships between the $n$-point functions which would point clearly to
a quantum mechanical origin for the fluctuations.
This is important because
it is unclear whether we could ever rule out a non-quantum
origin
(perhaps associated with new but non-inflationary
physics at early times)
using only the
two-point function.

Measurements of the CMB temperature anisotropy have
now reached sufficient accuracy that it is feasible to
estimate the three-point temperature autocorrelation function.
The most precise constraints come from
the Planck2013 dataset~\cite{Ade:2013ydc}.
But despite the quality of the measurements, the
signal-to-noise for any particular combination of
wavenumbers is still too low to allow the three-point function to
be reconstructed directly.
Instead, measurements are made by picking an
Ansatz or `template' for the way in which
the correlations change with wavenumber.
By comparing this template with
the CMB data over many different combinations of
wavenumber it is possible to attain
reasonable signal-to-noise.
This comparison carries a considerable
computational burden, so
constraints from the data are typically reported
as amplitudes for just a handful of well-known
templates, such as the `local', `equilateral'
and `orthogonal' shapes.
These amplitudes are often written
$\estfNLlocal$, $\estfNLequi$, $\estfNLortho$,
and so on.%
    \footnote{Here and throughout the remainder of the paper
    we distinguish quantities estimated from data
    by a hat.}

A specific inflationary model will be
characterized by a number $\Nparams$ of
adjustable parameters
$\lambda_i$, $1 \leq i \leq \Nparams$.
These may include
Lagrangian parameters which are
analogues of masses and couplings,
but in multiple-field models may also include
a specification of the initial conditions
in field-space.
To apply constraints
from $\estfNLlocal$, $\estfNLequi$, $\estfNLortho$, \ldots,
to such a model
its three-point function must be computed and
projected on to each of these templates.
This generates predictions
for each of the amplitudes
$\fNLlocal(\lambda_i)$,
$\fNLequi(\lambda_i)$,
$\fNLortho(\lambda_i)$,
\ldots.
The results obtained by a
microwave background survey
can then be converted into constraints on
the underlying parameters $\lambda_i$.

This approach is perfectly reasonable,
but there are reasons to
expect that it may not be optimal.
First, if the set of templates
does not cover the entire range of
three-point correlations which can be produced
by adjusting the parameters $\lambda_i$
then we are not making efficient use of the
data:
we should measure the amplitude of more templates
in order to obtain better constraints.
But, as many authors have pointed out,
it is not clear \emph{a priori} how large
a range of templates is required,
or how they should be chosen.

Second,
if our templates are chosen injudiciously
then there will come a point of diminishing returns
at which no new information is gained because
the shapes we are fitting are strongly
correlated with shapes
which have been tried before.
This is a reflection of a more general problem:
the error bars reported for any set of amplitudes
will typically be correlated,
with the correlation described by some covariance
matrix.
Without knowledge of these covariances we risk
underestimating the uncertainties associated
with our reconstruction of the parameters
$\lambda_i$.

In this paper we take a different approach.
We investigate the construction of
maximum-likelihood estimators
for the Lagrangian parameters $\lambda_i$
directly from the data.
(Because noise maps for the Planck2013
data release are not yet available,
we use the WMAP 9-year dataset.)
To decide which templates to use,
we catalogue the different types of
correlation which can be produced in
a well-specified class of models:
those whose fluctuations are described
the the effective field theory 
of inflation~\cite{Cheung:2007st}.
We construct the Fisher matrix associated
with these correlations and
use it to determine the principal
directions whose amplitudes can be measured
efficiently.
We account for the covariance between
measurements of these amplitudes and use
them to place constraints on
the underlying Lagrangian parameters.

\para{Summary}%
In~{\S\ref{sec:EFT}} we briefly review the effective field theory
approach to single-field inflation and catalogue the
operators arising from a general single-field action.
In~{\S\ref{sec:bispec}} we discuss the calculation of
bispectra corresponding to these operators,
and point out a number of subtleties which
must be borne in mind when interpreting our results.
In~{\S\ref{sec:estimating}} we assemble the formalism
which is used to extract constraints
from the CMB map:
in~{\S\ref{sec:how-many-shapes}}
we construct the Fisher matrix and use it
to determine the principal directions which
can be constrained efficiently,
and in~{\S\ref{sec:results}}
we report our measurements of their
amplitudes from the 9-year WMAP dataset.
{\S\ref{sec:model-constraints}} translates
these general constraints into the
language of specific models,
and~{\S\ref{sec:model-comparison}}
uses the framework of Bayesian model
comparison to gain some qualitative
information regarding the type of
model favoured by the data.
We conclude in~{\S\ref{sec:conclusions}}.
A short appendix tabulates the three-point
functions used in the main text.

\para{Notation}%
We use units in which $c = \hbar = 1$, and define the reduced
Planck mass $\Mp$ to be $\Mp^{-2} = 8 \pi G$.
Our index and summation conventions are
explained in the main text.

\section{Overview of the effective field theory of inflation}
\label{sec:EFT}
In this paper we focus on single-field models of inflation
which terminate in a unique minimum,
which we refer to as the `reheating minimum'.
In multiple-field models there are complications
associated with our freedom to set initial conditions.
These
determine the average field-space trajectory
followed by the region of the universe we choose
to study.
In a single-field model there is a unique trajectory which
terminates in the reheating minimum.

In both single- and multiple-field cases it is
quantum fluctuations around this average
field-space trajectory
which are inherited by the large-scale density perturbation,
but where there is no unique trajectory
the calculation of these fluctuations is a serious
computational challenge.
Their evolution must be followed until
an `adiabatic limit' has been reached, at which all
isocurvature modes become
exhausted~\cite{Weinberg:2003sw,Weinberg:2004kr,Weinberg:2004kf,
Meyers:2010rg,Elliston:2011dr}.
Normally this will require numerical methods.
In contrast,
the fluctuations produced in single-field inflation---%
or, more precisely, `single-clock' inflation---%
typically do not evolve
and can be computed analytically under certain circumstances.
Below,
we discuss the precise conditions which are required.

\para{Model parametrization}%
Our aim is to estimate the Lagrangian parameters which
characterize a single-field inflationary model.
How many such parameters are needed?
The answer depends on the range of behaviour which
we allow.
Cheung et al. gave an argument based on
nonlinearly realized Lorentz invariance
which, under certain conditions, constrains the possible
three-body interactions between scalar perturbations
on a smooth inflationary background~\cite{Cheung:2007st}.
This is the `effective field theory of inflation'.
In this section
we briefly review their construction.

The effective field theory is not used to
describe the background cosmology,
but only fluctuations around it.
Therefore
it is agnostic
regarding the precise mechanism of inflation.
The background
is assumed
to be described by a Robertson--Walker metric
\begin{equation}
    \d s^2 = - \d t^2 + a^2(t) \, \d \vect{x}^2 ,
\end{equation}
where $a(t)$ is the scale factor, $t$ is cosmic
time and $H(t) = \dot{a}/a$ is the
Hubble rate.
Since the background is evolving it spontaneously
breaks time-translation invariance (and therefore manifest
Lorentz invariance), but because the spatial slices are
homogeneous and isotropic the
background remains manifestly
invariant under spatial coordinate transformations.
We will use the terminology `coordinate transformations'
and `diffeomorphisms' interchangeably.

Knowledge of the background evolution is equivalent
to specifying $H(t)$ as a smooth function of $t$.
The condition that the universe is `single-clock'
is that a coordinate system exists in which only the
metric carries fluctuations;
in this coordinate system
all fields needed
to describe the matter sector are homogeneous, depending
only on the time $t$.
By analogy with similar constructions in
particle physics, Cheung et al. called this
coordinate system
the \emph{unitary gauge}. Where the matter sector is
described by a single scalar field $\phi$ it corresponds to
the gauge where fluctuations $\delta\phi$ vanish, but this is
not necessary.

To describe dynamics we require a Lagrangian.
A Lagrangian which is manifestly
invariant under the unbroken (linearly-realized)
group of purely spatial
coordinate transformations
will be a function $F$ which transforms as a scalar
under these diffeomorphisms.
Cheung et al. argued that
the most general such Lagrangian could be constructed as a
scalar function of the metric and the intrinsic and
extrinsic curvature tensors on the spatial slices, together
with their covariant derivatives~\cite{Cheung:2007st}. These may appear in
arbitrary combinations with
$t$ and the metric function
$g^{00}$, which are both invariant under spatial
coordinate transformations.
Therefore,
\begin{equation}
    S_{\text{gen}} = \int \d^4 x \; \sqrt{-g} \, F \big(
        R_{\mu\nu\rho\sigma}, K_{\mu\nu}, \grad_\mu, g^{00}, t
    \big) .
\end{equation}
By itself, this Lagrangian can describe fluctuations around
any cosmological background with linearly-realized spatial diffeomorphism
invariance. Specializing it to the background $H(t)$
fixes the background and linear terms,
\begin{equation}
    \label{eq:specialized-action}
    S = \int \d^4 x \; \sqrt{-g}
    \bigg(
        \frac{\Mp^2}{2} R + \Mp^2 \dot{H} g^{00} - \Mp^2 ( 3H^2 + \dot{H} )
        + \sum_{n \geq 2} F_n
        \big(
            \delta R_{\mu\nu\rho\sigma}, \delta K_{\mu\nu}, \grad_\mu, \delta g^{00}, t
        \big)
    \bigg) ,
\end{equation}
where $\delta R_{\mu\nu\rho\sigma}$
and $\delta K_{\mu\nu}$ are, respectively,
perturbations in the intrinsic and extrinsic curvature tensors, and
$\delta g^{00} = g^{00} + 1$ is the perturbation in the time--time metric
function or `lapse'.
The arbitrary functions $F_n$ are homogeneous polynomials of
order $n$, and therefore the leading correction to the first three
terms appearing in~\eqref{eq:specialized-action} is quadratic.

We have not yet made use of the requirement that
the full theory is invariant under time reparametrizations,
$t \rightarrow t' = t + \xi(\vect{x})$, where the
translation $\xi$ may be
a function of position.%
    \footnote{An arbitrary action of the form~\eqref{eq:specialized-action} can
    describe theories with this symmetry, in addition to others which
    do not.}
On an expanding cosmological background
this symmetry is spontaneously broken.
Nevertheless, once a choice
of spatially-invariant operators has been made
in Eq.~\eqref{eq:specialized-action},
the broken time-translation symmetry
is strong enough to fix
the interactions of one scalar mode.
To determine these interactions
we construct a new action by formally performing a time
translation $t \rightarrow t' = t - \pi$.
If we promote $\pi$ to a dynamical field
which shifts linearly under time translations
(that is, $\pi \rightarrow \pi' = \pi - \xi$
when $t \rightarrow t' = t + \xi$)
then the total action becomes manifestly invariant.
The field $\pi$ represents a scalar degree of freedom
in the system,
but its interactions are fixed uniquely by the
combination of tensors appearing in the $F_n$,
the background cosmology $H(t)$, and the
time translation
symmetry~\cite{Creminelli:2006xe,Cheung:2007st,Cheung:2007sv,
Weinberg:2008hq,Senatore:2009gt}

For this formalism to be useful it must be possible to
calculate each amplitude of interest using states
which contain no more than a handful of $\pi$ particles,
or $\pi$-lines in diagrammatic terms.
This is not generally true.
But if all background fields are time-independent then
rigid time translations
$t \mapsto t' = t + \xi$ (with $\xi$ a constant)
are a \emph{global} symmetry of the theory,
no matter what transformation law we ascribe to $\pi$.
Therefore $\pi$ must behave as a
Goldstone boson:
where it appears in the action
it must be accompanied by at
least one derivative.
In a process which takes place at a well-defined characteristic
energy scale $E$, each derivative will translate to a power of $E$.
The justification for neglecting diagrams which contain a large
number of $\pi$-lines is then the same as any effective field
theory of Goldstone modes, enabling a perturbative expansion
in powers of $E/M$ where $M$ is some large mass scale characterizing
the strength of the interactions.

For applications to inflation the background fields
are not constant but slowly varying, so rigid time translations
are only an approximate symmetry.
Therefore
terms involving undifferentiated powers of $\pi$ may
appear in the action,
although suppressed by dimensionless factors which
measure the degree to which the global symmetry is broken.
These generate effects
which are unaccompanied by powers of $E/M$
and therefore may be important at all energies.%
    \footnote{It is these terms which cause superhorizon
    evolution of the perturbations in multiple-field models.
    Their importance at all scales is reflected in the
    fact that they remain relevant even when $k/aH$ is very soft.}
However, provided the approximate symmetry is sufficiently good that
corrections to it are at least as small as the first
neglected power of $E/M$ it is still possible to carry out
a consistent calculation.
During inflation we are interested in the type of correlations
induced by each operator between modes of the quantized field
near the epoch of Hubble exit,
so the scale $E$ will be of order the Hubble
scale $H$.

At sufficiently high energies $E > \Edecouple$
the Goldstone mode decouples
from the remaining degrees of freedom in $\delta R_{\mu\nu\rho\sigma}$
and $\delta K_{\mu\nu}$.
(The notation `$\Edecouple$' was introduced by Cheung et al.~\cite{Cheung:2007st},
who emphasized that below $\Edecouple$ the mixing with gravitational
degrees of freedom cannot be ignored.)
If the decoupling scale $\Edecouple$
is at least modestly smaller than
$E = H$ then it is possible to study how each operator
generates correlations without including gravitational fluctuations.
In this paper we will work exclusively in the decoupling limit.
With this assumption,
Bartolo et al.~\cite{Bartolo:2010bj,Bartolo:2010di}
gave an effective action up to cubic terms,
\begin{equation}
\label{EFTLag}
\begin{split}
    \Seft = \int \d^4 x \; \sqrt{-g}
    \bigg\{
        &
        \Mp^2 \dot{H} (\partial_{\mu}\pi)^2 
        + 2 M_{2}^4
        \bigg[
            \dot{\pi}^2  - \dot{\pi} \frac{(\partial\pi)^2}{a^2}
        \bigg]
        - \frac{4}{3} M_{3}^4 \dot{\pi}^3
        \\ & \mbox{}
        - \frac{\bar{M}_{1}^3}{2 a^2}
        \bigg[
            {-2H(\partial\pi)^2}  + \frac{(\partial\pi)^2 \partial^{2}\pi}{a^2}
        \bigg]
        \\ & \mbox{} 
        - \frac{\bar{M}_{2}^2}{2 a^4}
        \Big[
            (\partial^{2}\pi)(\partial^{2}\pi) + H(\partial^{2}\pi)(\partial\pi)^2
            + 2 \dot{\pi}\partial^{2}\partial_{j}\pi\partial_{j}\pi 
        \Big]
        \\ & \mbox{}
        - \frac{\bar{M}_{3}^{2}}{2 a^4}
        \Big[
            (\partial^{2}\pi)(\partial^{2}\pi)
            + 2H \partial^{2}\pi (\partial\pi)^{2}
            + 2\dot{\pi}\partial^{2}\partial_{j}\pi\partial_{j} \pi
        \Big]
        \\ & \mbox{}
        - \frac{2 \bar{M}_{4}^3}{3 a^2} \dot{\pi}^{2}\partial^{2}\pi
        + \frac{\bar{M}_{5}^{2}}{3 a^4} \dot{\pi}(\partial^{2}\pi)^{2}
        + \frac{\bar{M}_{6}^{2}}{3 a^4} \dot{\pi}(\partial_{i}\partial_{j}\pi)^{2}
        - \frac{\bar{M}_{7}}{3! \cdot a^6} (\partial^{2}\pi)^{3}
        \\ & \mbox{}
        - \frac{\bar{M}_{8}}{3! \cdot a^6} \partial^{2}\pi(\partial_{i}\partial_{j}\pi)^2
        - \frac{\bar{M}_{9}}{3! \cdot a^6} \partial_{i}\partial_{j}\pi \partial_{j}\partial_{k}\pi \partial_{k}\partial_{i}\pi
    \bigg\} .
\end{split}
\end{equation}
Our notation has been chosen to match
Refs.~\cite{Bartolo:2010bj,Bartolo:2010di}.
The mass scales $M_2$, $M_3$ and $\bar{M}_1, \ldots, \bar{M}_9$
characterize the model under
consideration.%
    \footnote{To aid intuition, the powers of the
    $M_i$ and $\bar{M}_i$ appearing in Eq.~\eqref{EFTLag}
    have been chosen so that the $M_i$ and $\bar{M}_i$
    all have dimensions of mass when using natural units
    in which $c = \hbar = 1$.
    In some cases this means that positive integer powers of
    masses appear, such as $M_3^4$, which can only
    be positive if $M_3$ is real.
    In reality there is an undetermined sign which we
    are suppressing, so that $M_3^4$ should be regarded
    as an object which can be either positive or negative.
    The associated mass scale is $|M_3^4|^{1/4}$.\label{footnote:mass-signs}}
Terms decorated with a bar are associated
with operators involving the extrinsic curvature $\delta K_{\mu\nu}$,
whereas
unbarred terms correspond to powers of $\delta g^{00}$.
In writing Eq.~\eqref{EFTLag}, Bartolo et al. did not
include all possible operators:
they neglected higher-derivative operators
containing derivatives of the form
$\grad_\mu \delta g^{00}$
and $\grad_\lambda K_{\mu\nu}$,
and from the lowest-derivative combinations
for each $M_i$ or $\bar{M}_i$ they
retained only
terms which gave a parametrically large contribution
to the three-point function.
We can expect the higher-derivative operators to be small
provided the mass scales $M_i$, $\bar{M}_i$ are sufficiently large,
which is already the condition that the EFT is predictive.
Therefore, although~\eqref{EFTLag} does not represent the most general
set of interactions, it is reasonable to speculate that it may approximate
the most general set of \emph{observable} interactions
for a smooth background $H(t)$.
In this paper we only consider backgrounds
which satisfy this smoothness requirement.
The properties of fluctuations over backgrounds
which are not sufficiently smooth
require a separate analysis; for example,
see Refs.~\cite{Bartolo:2013exa,Adshead:2013zfa}.

When is the decoupling approximation valid?
Estimates for the scale $\Edecouple$ were
given by Cheung et al.~\cite{Cheung:2007st},
but strictly this scale can be determined
only when the $M_i$ and $\bar{M}_i$
are known
and therefore it must be checked \emph{a posteriori}.
As an example,
in canonical single-field inflation, Cheung et al.
argued that $\Edecouple \sim \epsilon^{1/2} H$,
where $\epsilon \equiv - \dot{H}/H^2$ is a measure of the degree to
which
the global symmetry of
rigid time translations is broken.
If $\epsilon \ll 1$ then a decoupling regime can exist near the
Hubble scale.

The scales $M_i$ and $\bar{M}_i$ can
be adjusted to reproduce the results
of well-known models including canonical single-field inflation,
Dirac--Born--Infeld inflation~\cite{Alishahiha:2004eh}
and Ghost Inflation \cite{ArkaniHamed:2003uz}.
Alternatively they may be allowed to float.
The action~\eqref{EFTLag} then explores
a range of interactions for fluctuations
on a quasi-de Sitter background
with nonlinearly realized Lorentz invariance,
subject to the proviso (as described above) that only the
dominant term for each $M_i$ and $\bar{M}_i$ has been retained.
In principle these mass
scales depend on time, but because we are taking
the time-dependence of background quantities to be very weak
we will treat them as constants.

\section{Calculation of the bispectrum}
\label{sec:bispec}
In this paper our aim is to estimate the parameters
$M_i$, $\bar{M}_i$ by using observations to indirectly
reconstruct the two- and three-point functions
$\langle \pi \pi \rangle$
and
$\langle \pi \pi \pi \rangle$.
By itself,  $\pi$ is not an observable
and neither are its correlations:
the measurable quantity is the temperature fluctuation $\delta T/T$
as a function of angular position on the sky.
Typically this is decomposed into harmonics, generating corresponding
amplitudes $a_{\ell m}$,
\begin{equation}
    \frac{\delta T(\hat{\vect{n}})}{T} =
    \sum_{\ell m} a_{\ell m} Y_{\ell m}(\hat{\vect{n}}) ,
\end{equation}
where $\hat{\vect{n}}$ represents an orientation on the sky
and
$Y_{\ell m}(\hat{\vect{n}})$ is a conventionally-normalized
spherical harmonic.
The amplitude $a_{\ell m}$ can be predicted in terms of
primordial quantities using the formula%
    \footnote{We have absorbed a conventional
    factor of $3/5$ into the normalization of the
    transfer function.}
\begin{equation}
    \label{eq:deltaT-alm}
    a_{\ell m} = 4\pi(-\im)^\ell
    \int \frac{\d^3 k}{(2\pi)^3} \Delta_\ell(k) \zeta(\vect{k})
    Y_{\ell m}(\hat{\vect{k}}) ,
\end{equation}
where
the `curvature perturbation' $\zeta = \delta \ln a(\vect{x},t)$ represents
a fluctuation in the local scale factor $a(\vect{x},t)$.
It can be related to $\pi$ via $\zeta = - H \pi$
up to terms which vanish in the limit $k/aH \rightarrow 0$,
where $k$ is the Fourier mode under consideration
and $aH$ is the comoving wavenumber associated with the Hubble length.

In writing~\eqref{eq:deltaT-alm} we have assumed that,
for each relevant Fourier mode, $\zeta(\vect{k})$ attains
a practically time-independent value by some time during the
radiation era.
The transfer function $\Delta_\ell(k)$
describes the subsequent
process by which this time-independent seed
perturbation
is taken up by fluctuations in the primordial plasma
and propagated to the surface of last scattering, where it
constitutes a temperature fluctuation $\delta T$.
Under these circumstances,
Eq.~\eqref{eq:deltaT-alm} shows that the $n$-point functions
of the $a_{\ell m}$ can be linearly related to the $n$-point functions
of $\zeta(\vect{k})$, and therefore $\pi(\vect{k})$,
provided we evaluate the curvature perturbation in~\eqref{eq:deltaT-alm}
at a time when the $\Or(k/aH)$ corrections in the relationship between
$\pi$ and $\zeta$ are negligible.

\para{Correlation functions of $\zeta$}%
Therefore, we must estimate the
correlation functions of $\zeta$ at the time they
achieve their constant values.
It is this requirement which makes
the study of multiple-field models
challenging~\cite{Elliston:2011dr,Dias:2012qy},
because it is difficult to predict in advance
when the time-independent epoch will occur.
In single-field models
the situation is simpler
because the approximate global symmetry
under rigid time translations
(together with certain technical assumptions)
is sufficient
to prove the
operator statement $\dot{\zeta} = 0$
in the limit
$k/aH \rightarrow 0$~\cite{Assassi:2012et,Senatore:2009cf,Senatore:2010jy,
Senatore:2010wk,Baumann:2011su,Baumann:2011dt}.
Therefore all correlation functions of $\zeta$ are constant
on superhorizon scales, where $|k/aH|$ is negligible.
An important consequence of this result is
that subleading terms in the
effective action~\eqref{EFTLag}
map to subleading terms in each $n$-point function~\cite{Dias:2012qy},
so to obtain
a lowest-order
result there is no need to consider corrections to~\eqref{EFTLag}
due to our neglect of time dependence in the $M_i$, $\bar{M}_i$.

In perturbation theory, a three- or higher $n$-point function
is computed by integrating the reaction rate for an $n$-body
interaction together with factors representing the available interaction
volume and the probability for suitable particles to be present.
These techniques
were first applied to inflation by Maldacena~\cite{Maldacena:2002vr}
and later refined by various authors~\cite{Creminelli:2003iq,Alishahiha:2004eh,
Seery:2005wm,Seery:2005gb,Weinberg:2005vy,Chen:2006nt,Burrage:2011hd,Elliston:2012ab}.
We refer to this literature for technical details.
In this section we wish to emphasize that, in the context of a general
effective field theory, there are subtleties associated with
computation of the field mode functions. These represent the
amplitude
for single-particle excitations of the vacuum.
Therefore their properties significantly
influence
the $n$-point functions
because they determine
the probability for
particles to be
present in the interaction region.

Bartolo et al. noted that
the scales
$M_1$, $\bar{M}_1$, $\bar{M}_2$ and $\bar{M}_3$
in Eq.~\eqref{EFTLag}
are correlated with
contributions to the second-order effective action,
and
of these $\bar{M}_2$ and $\bar{M}_3$
generate kinetic terms involving fourth-order
derivatives.
Kinetic terms of this type
had previously been encountered in the `Ghost Inflation'
scenario proposed by Arkani-Hamed et al.~\cite{ArkaniHamed:2003uz}.
Such terms are problematic because they
imply that the mode functions can no longer be
expressed in terms of elementary functions.
This obstructs analytic integration
of the interaction rate
and hence each $n$-point function.
In scenarios which require these high-order
kinetic terms,
exact
results for the correlation functions
typically require numerical calculation.

Bartolo et al. gave an explicit formula
for the mode functions including
the contribution of fourth-order terms,
expressed
in terms of hypergeometric functions and generalized
Laguerre polynomials~\cite{Bartolo:2010bj},
and performed an analysis 
of its influence on each $n$-point function~\cite{Bartolo:2010bj,Bartolo:2010di}.
They concluded that the fourth-order terms
could significantly modify propagation
deep within the horizon,
but produced qualitatively similar results
near the epoch of horizon exit.
Since the correlations we are seeking to study
are exponentially dominated by interactions
occurring near this epoch,
this implies that an acceptable estimate of
the bispectrum \emph{shape} can be obtained
using a simpler mode function which does not
account for fourth-order
contributions.
The penalty for this approximation
is an uncertainty in the amplitude,
which arises from a difference in
\emph{normalization} between
the mode functions with and without
the inclusion of fourth-order
terms.
For more details we refer to the
discussion in Refs.~\cite{Bartolo:2010bj,Bartolo:2010di}.

In this paper we follow
Bartolo et al. and estimate each bispectrum shape
by neglecting the influence of fourth-order terms.
This means that our results must be interpreted
with some care:
\begin{enumerate}
    \item When applied to a model for which
    $\bar{M}_2 = \bar{M}_3 = 0$, our results are
    exact within the approximations which have
    already been discussed.
    In this case, we expect both our qualitative
    and quantitative conclusions to be reliable.

    \item When applied to a model for which
    at least one of
    $\bar{M}_2$ or $\bar{M}_3$ is nonzero,
    the normalization of our bispectra will
    be incorrect for the reasons just explained.
    This uncertainty in normalization affects
    the bispectrum for each operator in Eq.~\eqref{EFTLag},
    not just those associated with the scales
    $\bar{M}_2$ and $\bar{M}_3$---%
    but we expect
    that it should be approximately the same for all of them.
    In this scenario, our quantitative estimates for
    the mass scales $M_i$, $\bar{M}_i$ are not reliable.
    However, qualitative conclusions regarding the
    relative importance of each operator should be
    unaffected because
    ratios of these mass scales divide out any
    uncertainty in normalization.
    
    To obtain reliable quantitative estimates of the
    mass scales when at least one of
    $\bar{M}_2$ or $\bar{M}_3$ is nonzero,
    it would be necessary to substitute
    numerical calculations of the bispectra
    in our analysis.
    In addition, the likelihood
    function to be discussed in~{\S\ref{sec:estimating}}
    would no longer be approximately Gaussian
    and the analysis to follow
    should be replaced by a more sophisticated
    numerical
    exploration of the likelihood surface.
    
    These modifications significantly increase
    the complexity of the analysis. They would
    certainly
    be required if observations provided pressure
    to include an $\bar{M}_2$ or $\bar{M}_3$ term
    in the effective Lagrangian.
    At present, our view is that such a step in
    complexity is not justified by the
    data.
\end{enumerate}

\section{Estimating the EFT mass scales}
\label{sec:estimating}

\para{Bispectrum of curvature perturbation}%
Under the approximations discussed in~{\S\ref{sec:bispec}},
the shapes of the bispectra generated by each operator
in Eq.~\eqref{EFTLag}
were plotted in Ref.~\cite{Bartolo:2010bj}.
We tabulate analytical results
for the corresponding three-point functions
(which were not given explicitly in Ref.~\cite{Bartolo:2010bj})
in Appendix~\ref{sec:appendbispec}.
The total three-point function for $\zeta$ should be obtained
by summing these contributions,
weighted by an appropriate mass scale
$M_i$ or $\bar{M}_i$.

In what follows it will be convenient to collect these mass
scales, together with other normalization
factors, into dimensionless combinations
$\lambda_\alpha$
given in Table~\ref{tab:coeff}.
There are eleven independent mass scales and therefore
eleven independent $\lambda_\alpha$.
We use Greek indices $\alpha$, $\beta$, \ldots,
to label these scales and the corresponding
Lagrangian operators,
which we write abstractly as $\Op^\alpha$.
Each index ranges over the values
$A$, $B$, \ldots, $K$,
and the effective action
is the combination $\Seft = \int \d^4 x \, \sqrt{-g}
\sum_\alpha \lambda_\alpha \Op^\alpha$.
We position indices so that the normal
rules of the
Einstein summation convention are respected,
but
for clarity we will usually write
summations over these indices explicitly.
\begin{table}
\centering
    \small
	\heavyrulewidth=.08em
	\lightrulewidth=.05em
	\cmidrulewidth=.03em
	\belowrulesep=.65ex
	\belowbottomsep=0pt
	\aboverulesep=.4ex
	\abovetopsep=0pt
	\cmidrulesep=\doublerulesep
	\cmidrulekern=.5em
	\defaultaddspace=.5em
	\renewcommand{\arraystretch}{2.4}


\begin{tabular}{svv}
    \toprule
    \multicolumn{1}{l}{\textbf{Parameter}}
        & \multicolumn{2}{c}{\textbf{expressed as a mass scale}} \tabularnewline
        
    & \multicolumn{1}{c}{in terms of $H$} & \multicolumn{1}{c}{$H$ eliminated}
        \\ \cmidrule{2-3}
        
 	\lambda_{A}
 	    & -\frac{65}{20736} \frac{1}{\pi^4 \epsilon^3 \cs^4 \As^{2}} \frac{\bar{M}_{1}^{3} H^3}{\Mp^6}
 	    & -\frac{65}{648 \sqrt{2}} \frac{1}{\pi \epsilon^{3/2} \cs^{5/2} \As^{1/2}} \frac{\bar{M}_{1}^{3}}{\Mp^3} \tabularnewline
	
	\lambda_{B}
	    & -\frac{85}{10368} \frac{1}{\pi^4 \epsilon^3 \cs^2 \As^{2}} \frac{M_{2}^{4} H^2}{\Mp^6}
	    & -\frac{85}{1296} \frac{1}{\pi^2 \epsilon^{2} \cs \As} \frac{M_{2}^{4}}{\Mp^{4}}  \tabularnewline
	
	\lambda_{C}
	    & -\frac{325}{62208} \frac{1}{\pi^4 \epsilon^3 \cs^4 \As^{2}} \frac{\bar{M}_{2}^{2} H^4}{\Mp^{6}}
	    & -\frac{325}{972} \frac{1}{ \epsilon \cs^{2}} \frac{\bar{M}_{2}^{2}}{\Mp^2} \tabularnewline
	
	\lambda_{D}
	    & \frac{5}{3888} \frac{1}{\pi^4 \epsilon^3 \As^{2}} \frac{M_{3}^{4} H^2}{\Mp^{6}}
	    & \frac{5}{486} \frac{\cs}{\pi^2 \epsilon^{2} \As} \frac{M_{3}^{4}}{\Mp^{4}} \tabularnewline
	
	\lambda_{E}
	    & -\frac{65}{7776} \frac{1}{\epsilon^3 \cs^4 \As^{2}} \frac{\bar{M}_{3}^{2} H^4}{\Mp^{6}}
	    & -\frac{130}{243} \frac{1}{\epsilon \cs^{2}} \frac{\bar{M}_{3}^{2}}{\Mp^{2}}  \tabularnewline
		
	\lambda_{F}
	    & \frac{5}{3888} \frac{1}{\pi^4 \epsilon^3 \cs^2 \As^{2}} \frac{\bar{M}_{4}^{3} H^3}{\Mp^{6}}
	    & \frac{5 \sqrt{2}}{243} \frac{1}{\pi \epsilon^{3/2}\cs^{1/2} \As^{1/2}} \frac{\bar{M}_{4}^{3}}{\Mp^{3}} \tabularnewline
	
	\lambda_{G}
	    &- \frac{65}{46656} \frac{1}{\pi^4 \epsilon^3 \cs^4 \As^{2}} \frac{\bar{M}_{5}^{2} H^4}{\Mp^{6}}
	    & -\frac{65}{729} \frac{1}{\epsilon \cs^{2}} \frac{\bar{M}_{5}^{2}}{\Mp^{2}} \tabularnewline
	
	\lambda_{H}
	    & -\frac{65}{186624} \frac{1}{\pi^4 \epsilon^3 \cs^4 \As^{2}} \frac{\bar{M}_{6}^{2} H^4}{\Mp^{6}}
	    & -\frac{65 }{2916} \frac{1}{\epsilon \cs^{2}} \frac{\bar{M}_{6}^{2}}{\Mp^{2}} \tabularnewline
	
	\lambda_{I}
	    & \frac{115}{69984} \frac{}{\pi^4 \epsilon^3 \cs^6 \As^{2}} \frac{\bar{M}_{7} H^5}{\Mp^{6}}
	    &  \frac{460 \sqrt{2}}{2187} \frac{\pi \As^{1/2}} {\epsilon^{1/2} \cs^{7/2}} \frac{\bar{M}_{7}}{\Mp} \tabularnewline
	
	\lambda_{J}
	    & \frac{115}{279936} \frac{1}{\pi^4 \epsilon^3 \cs^6 \As^{2}} \frac{\bar{M}_{8} H^5}{\Mp^{6}}
	    &  \frac{115 \sqrt{2}}{2187} \frac{\pi \As^{1/2}}{\epsilon^{1/2} \cs^{7/2}} \frac{\bar{M}_{8}}{\Mp} \tabularnewline
	
	\lambda_{K}
	    & -\frac{115}{559872} \frac{1}{\pi^4 \epsilon^3 \cs^6 \As^{2}} \frac{\bar{M}_{9} H^5}{\Mp^{6}} 
	    & -\frac{115}{2187 \sqrt{2}} \frac{\pi \As^{1/2}}{\epsilon^{1/2} \cs^{7/2}} \frac{\bar{M}_9}{\Mp} \\
	\bottomrule
\end{tabular}
\caption{\label{tab:coeff}Parameters $\lambda_{\alpha}$ in terms of the coefficients in the Lagrangian}
\end{table}
With these choices, we find
\begin{equation}
    \label{eq:zeta-bispectrum}
    B_\zeta(k_1, k_2, k_3) = \frac{3}{5} 
        \sum_\alpha \lambda_\alpha B^\alpha (k_1, k_2, k_3) ,
\end{equation}
where $B$ labels the bispectrum,
defined
so that (for example)
\begin{equation}
    \langle \zeta(\vect{k}_1) \zeta(\vect{k}_2) \zeta(\vect{k}_3) \rangle
    = (2\pi)^3 \delta(\vect{k}_1 + \vect{k}_2 + \vect{k}_3) B_\zeta(k_1, k_2, k_3) ,
\end{equation}
and similarly for the $\pi$ three-point function,
which produces a bispectrum
$B^\alpha$ for each operator $\Op^\alpha$.
In Eq.~\eqref{eq:zeta-bispectrum}
the normalization of each $\lambda_\alpha$ has been adjusted so that
the $B^\alpha$ satisfy
\begin{equation}
    \label{eq:bispectrum-normalization}
    \frac{B^\alpha(k,k,k)}{6{P}_\zeta(k)^2} = 1 ,
\end{equation}
where ${P}_\zeta(k)=2\pi^2\As/k^3$
is the power spectrum
and $\As$ is the scalar amplitude.
Each bispectrum is evaluated at the equilateral point
and in principle depends on the side length $k$.
However, because the
$n$-point functions
we study are nearly scale invariant
(which for the bispectra implies
$B^\alpha(k,k,k) \sim k^{-6}$),
the precise choice of
scale used to fix this normalization is
unimportant.
For a
precisely local bispectrum,
our convention~\eqref{eq:bispectrum-normalization}
would make the corresponding $\lambda_\alpha$ equal to the conventional
nonlinearity parameter $\fNLlocal$.
In general, however, the $B^\alpha$ will not be local
and
although each nonlinearity parameter such as $\fNLlocal$, $\fNLequi$, etc.,
will be a linear combination of the $\lambda_\alpha$, the coefficients
in these combinations
need not be simple.

\para{Projection to the CMB bispectrum}%
Eq.~\eqref{eq:deltaT-alm} shows that measurements of the microwave
background anisotropies do not
furnish information about $B_\zeta$ directly,
but only via correlation
functions of the $a_{\ell m}$.
The first such correlation function
which contains accessible information
regarding $B_\zeta$ is the
three-point function
$\langle a_{\ell_1 m_1} a_{\ell_2 m_2} a_{\ell_3 m_3} \rangle$.
It is conventional to extract a combinatorical factor
$\mathcal{G}^{l_1 l_2 l_3}_{m_1 m_2 m_3}$---the
so-called `Gaunt integral'---%
which is nonzero only for allowed combinations of the $\ell_i$ and $m_i$.
The remainder of the correlation function is written as
a `reduced bispectrum' $b_{\ell_1 \ell_2 \ell_3}$,
\begin{equation}
    \langle a_{\ell_1 m_1} a_{\ell_2 m_2} a_{\ell_3 m_3} \rangle =
    b_{\ell_1 \ell_2 \ell_3}
    \mathcal{G}^{\ell_1 \ell_2 \ell_3}_{m_1 m_2 m_3} .
\end{equation}

Our task is to determine $b_{\ell_1 \ell_2 \ell_3}$ given $B_\zeta$.
A strategy for doing so was developed by Fergusson, Shellard and
collaborators~\cite{Fergusson:2009nv,Fergusson:2010dm,Fergusson:2010ia,Fergusson:2010gn,Regan:2010cn}
and extended by other authors~\cite{Byun:2013jba,Battefeld:2011ut}.
We briefly recount the steps in this strategy, using the notation of
Refs.~\cite{Regan:2013wwa,Regan:2013jua}.
First, for each bispectrum $B^\alpha$ one defines a corresponding
dimensionless
`shape function' $S^\alpha$ using a fixed reference bispectrum
$\Bref$,
\begin{equation}
    S^\alpha(k_1, k_2, k_3) \equiv \frac{B^\alpha(k_1, k_2, k_3)}{\Bref(k_1, k_2, k_3)} .
\end{equation}
In principle, our final predictions do not depend on the choice of $\Bref$.
In practice we will be forced to make approximations,
some of which may introduce a residual dependence on $\Bref$.
For this reason
it is helpful to choose a form which has good numerical
properties;
often it is a good choice to fix a $\Bref$ which shares
features similar to the $B^\alpha$. In this paper we will use
the `constant' bispectrum~\cite{Fergusson:2008ra},
\begin{equation}
    \Bref(k_1, k_2, k_3) =
    6\left( \frac{2\pi^2 \As}{k_1 k_2 k_3} \right)^2 .
\end{equation}

Second, one chooses a set of functions $\Rmode^n$
which furnish at least an approximate basis
for the functions $S^\alpha$.
We define coefficients $\alpha^\alpha_n$ so that
\begin{equation}
    \label{eq:shape-decomposition}
    S^\alpha (k_1, k_2, k_3) \approx
    \sum_n \alpha^\alpha_n \Rmode^n(k_1, k_2, k_3) .
\end{equation}
In practice it is only possible to retain a finite number
of the $\Rmode^n$,%
    \footnote{The error associated with this truncation
    is one place where residual dependence on the
    reference bispectrum $\Bref$ can appear.}
so they should be chosen to give
an acceptable approximation for each $S^\alpha$ using only
a small number of modes.
For details regarding the
construction of suitable $\Rmode^n$
we refer to the literature~\cite{Fergusson:2010gn,Byun:2013jba,Battefeld:2011ut}.
It follows that, to a good approximation, the
$\zeta$ bispectrum can be written
\begin{equation}
    B_\zeta(k_1, k_2, k_3) \approx
    \frac{3}{5} \Bref(k_1, k_2, k_3)
    \sum_n \sum_\alpha \lambda_\alpha \alpha^\alpha_n \Rmode^n(k_1, k_2, k_3) .
\end{equation}

The map from $\zeta(\vect{k})$ to $a_{\ell m}$
expressed by Eq.~\eqref{eq:deltaT-alm} is linear,
and therefore
the observable quantity
$b_{\ell_1 \ell_2 \ell_3}$ must be
proportional to a linear combination of the
coefficients $\sum_\alpha \lambda_\alpha \alpha^\alpha_n$.
Therefore we can write
\begin{equation}
    b_{\ell_1 \ell_2 \ell_3} =
    \sum_{n, m} {\Gamma_n}^m b^{n}_{\ell_1 \ell_2 \ell_3}
    \sum_{\alpha} \lambda_\alpha \alpha^\alpha_m
    =
    \sum_{\alpha} \lambda_\alpha b_{\ell_1 \ell_2 \ell_3}^{\alpha},
\end{equation}
where
$b^\alpha_{\ell_1 \ell_2 \ell_3}$ is the reduced angular bispectrum
associated with the operator $\Op^\alpha$,
\begin{equation}
    b^\alpha_{\ell_1 \ell_2 \ell_3} \equiv
    \sum_{n,m} \alpha^\alpha_m {\Gamma_n}^m b^n_{\ell_1 \ell_2 \ell_3} ,
\end{equation}
and
the basis functions
$b^{n}_{\ell_1 \ell_2 \ell_3}$
are defined in Refs.~\cite{Fergusson:2009nv,Regan:2013wwa}.
They do not depend on any details of the cosmological model,
which is carried only by the `transfer matrix' ${\Gamma_n}^m$.
This can be expressed as an integral over the linear transfer
function $\Delta_\ell(k)$.
The virtue of the
approach of Fergusson, Shellard et al. is that
calculation of
${\Gamma_n}^m$
is numerically more tractable than calculation of
an arbitrary bispectrum.
Explicit formulae for the
$b_{\ell_1 \ell_2 \ell_3}^{n}$ and ${\Gamma_n}^m$ were given in
Refs.~\cite{Regan:2013wwa,Regan:2013jua}.
To compress notation we define
$\bar{\alpha}^\alpha_n \equiv \sum_m {\Gamma_n}^m \alpha^\alpha_m$
and $\bar{\beta}_n = \sum_\alpha \lambda_\alpha \bar{\alpha}^\alpha_n$,
from which it follows that
\begin{equation}
    \label{eq:angular-bispectrum-decomposition}
    b_{\ell_1 \ell_2 \ell_3} \approx
        \sum_n \bar{\beta}_n b_{\ell_1 \ell_2 \ell_3}^{n} .
\end{equation}

This projection procedure introduces correlations between
the observable bispectra
$b^\alpha_{\ell_1 \ell_2 \ell_3}$
produced by different Lagrangian operators,
even if the corresponding
primordial bispectra $B^\alpha(k_1, k_2, k_3)$
are nearly uncorrelated.
\label{page:late-correlation}
We will return to this issue in~{\S\ref{sec:how-many-shapes}} below.

\para{Comparison with data}%
It follows from Eq.~\eqref{eq:angular-bispectrum-decomposition}
that
information about the observable bispectrum from
a microwave background survey can be reduced to estimates of the
$\bar{\beta}_n$ and their covariances.
We denote these estimates
$\hat{\beta}_n$
and write their covariance matrix
$\Cov_{mn}$,
\begin{equation}
    \label{eq:covariance-def}
    \Cov_{mn} \approx \langle \Delta \hat{\beta}_m \Delta \hat{\beta}_n \rangle ,
\end{equation}
where $\Delta \hat{\beta}_n$ is the deviation of the observed
$\hat{\beta}_n$ from its expected value,
$\Delta \hat{\beta}_n \equiv \hat{\beta}_n - \bar{\beta}_n$.
The standard methods of linear algebra
can be used to obtain an orthonormal combination of
bispectra from a Cholesky decomposition of this
matrix~\cite{Fergusson:2009nv,Fergusson:2010dm,
Fergusson:2010ia,Fergusson:2010gn,Regan:2010cn,
Regan:2013wwa,Regan:2013jua}.
In the interests of simplicity we assume this has been done,
which makes $\Cov_{mn}$
equal (for the rotated bispectra) to the identity matrix.%
    \footnote{We estimate $\Cov_{mn}$
    from the covariance matrix of the cubic needlet statistic,
    after changing basis to the $\Rmode_n$
    as described in {\S}II.B of Ref.~\cite{Regan:2013wwa}.
    (See also Ref.~\cite{Regan:2013jua}.)
    The signal-to-noise for the bispectrum~\eqref{eq:covariance-def}
    is roughly
    \begin{equation*}
        \Big(
            \frac{\text{S}}{\text{N}}
        \Big)^2
        \approx
        -2 \ln \Lik
        \approx
        \sum_{mn}
            \bar{\beta}_m
            (\Cov^{-1})^{mn}
            \bar{\beta}_n ,
    \end{equation*}
    making $\Cov_{mn}$ a Fisher estimate of the covariance
    for the $\bar{\beta}_n$.
    Under the assumption that the bispectrum is small
    we assume that this covariance matrix is
    a reasonable approximation to
    $\langle \Delta \beta_m \Delta \beta_n \rangle$.

    To compute $\Cov_{mn}$
    we use a suite of $50,000$ Gaussian simulations,
    and for the change-of-basis coefficients we use
    a suite of $1,000$ non-Gaussian simulations.
    These simulations incorporate the effect of the WMAP beam
    and mask for each channel, including noise with
    variance-per-pixel determined by the WMAP 9-year
    data release.
    In the rotated basis, where we choose
    $\Cov_{mn} = \delta_{mn}$, 
    all these details are transferred to the definition of
    the $\hat{\beta}_n$.
    \label{footnote:wmap-monte-carlo}}
    
For a set of measurements $\hat{\beta}_n$, the
likelihood function $\Lik$
represents the probability
that these values would be observed given a
particular model for their origin---in this case,
the effective
Lagrangian~\eqref{EFTLag} with parameters $\lambda_\alpha$.
Assuming that the $\hat{\beta}_n$ are Gaussian distributed, this
probability can be written
\begin{equation}
    \label{eq:likelihood}
    \Lik( \hat{\beta}_n | \lambda_\alpha )
    =
    \frac{1}{\sqrt{2\pi \det \Cov}}
    \exp\bigg(
        - \frac{1}{2} \sum_{m, n} (\Cov^{-1})^{mn}
            \Delta \hat{\beta}_m
            \Delta \hat{\beta}_n
    \bigg) .
\end{equation}

\para{Maximum likelihood estimator}%
It is now simple to construct a
maximum likelihood
estimator for the $\lambda_\alpha$
by finding the combination 
which has the greatest likelihood
given the data.
This gives the estimate
\begin{equation}
    \hat{\lambda}_\alpha
    =
    \sum_\beta \hat{b}^\beta (\Fisher^{-1})_{\beta\alpha} ,
\end{equation}
where $\hat{b}^\alpha$ is defined by
\begin{equation}
    \hat{b}^\alpha = \sum_{m,n} \hat{\beta}_m (\Cov^{-1})^{mn}
    \bar{\alpha}^\alpha_n .
\end{equation}
The matrix $\vect{\Fisher}$ is the Fisher
matrix associated with the likelihood~\eqref{eq:likelihood},
\begin{equation}
    \label{eq:fisher-def}
    \Fisher^{\alpha\beta} =
        - \frac{\partial^2 \ln \Lik}{\partial \lambda_\alpha \partial \lambda_\beta}
        =
        \sum_{m,n}
        \bar{\alpha}^\alpha_m
        (\Cov^{-1})^{mn}
        \bar{\alpha}^\beta_n .
\end{equation}
Truncating at the quadratic level,
its inverse is formally the covariance matrix of the $\hat{\lambda}_\alpha$,
\begin{equation}
    \label{eq:mle-covariance}
    \langle (\hat{\lambda}_\alpha-{\lambda}_\alpha) (\hat{\lambda}_\beta-{\lambda}_\beta) \rangle
    = (\Fisher^{-1})_{\alpha\beta} .
\end{equation}
In the rotated basis, where $\Cov_{mn}$
is defined to be the unit matrix,
Eq.~\eqref{eq:fisher-def}
makes $\Fisher^{\alpha\beta}$ the square of the matrix
$\bar{\alpha}^\alpha_n$ which expresses the decomposition of the
angular bispectrum corresponding to the operator $\Op^\alpha$.
In practice the Fisher formalism
and Eq.~\eqref{eq:mle-covariance}
are likely to be trustworthy only
in the limit of sufficiently high signal-to-noise.

The maximum likelihood estimator
is an essentially frequentist concept,
as is its variance~\eqref{eq:mle-covariance}.
In a Bayesian framework one should instead interpret
$(\Fisher^{-1})_{\alpha\beta}$ as the covariance
of the posterior probability distribution
of the parameters $\lambda_\alpha$,
constructed from a single set of measurements
$\hat{\beta}_n$,
assuming that any prior probabilities for
the $\lambda_\alpha$ are flat over the range of
interest.

\subsection{How many independent shapes?}
\label{sec:how-many-shapes}
This analysis applies provided the matrix
$\Fisher^{\alpha\beta}$ is invertible.
However, invertibility
may fail if two linear combinations of the operators
$\Op^\alpha$ produce
nearly degenerate angular bispectra.
This would imply that $\Fisher^{\alpha\beta}$ has
an approximate null eigenvector.

The appearance of exact
or approximate null eigenvectors implies that
the likelihood function is a singular Gaussian distribution:
it does not vary along directions in parameter space
which correspond to the null eigenvectors.
Therefore the variance of the maximum likelihood
estimator~\eqref{eq:mle-covariance} is formally
infinite for all $\hat{\lambda}_\alpha$.
To deal with this one should first discard those combinations
of parameters which are unconstrained by the likelihood
function.
This is necessary even in the case of an
approximate null eigenvector,
because
although the Fisher matrix may be formally invertible
it will usually be ill-conditioned.
Therefore we should trust
a numerical inversion only if
it is possible to compute
$\Fisher^{\alpha\beta}$
to very high accuracy.
Typically this cannot be done because
the accuracy with which we know
$\Fisher^{\alpha\beta}$ is limited by our ability to estimate
$\Cov_{mn}$, and by the numerical integrations required
to compute
$\bar{\alpha}_n^\alpha$.
For a discussion of the issues involved in handling
singular Fisher matrices, see (for example) Ref.~\cite{Vallisneri:2007ev}.


\para{Measures of correlation}%
Two operators will produce degenerate bispectra if their
decomposition coefficients $\alpha^\alpha_m$
or $\bar{\alpha}^\alpha_m$
are nearly the same.
These two measures do not have to agree,
because
(as explained on p.~\pageref{page:late-correlation})
the projection
from $\alpha^\alpha_m$ to $\bar{\alpha}^\alpha_m$
can change the degree of correlation.
The Fisher matrix $\vect{\Fisher}$ is constructed from
angular bispectra,
and therefore---as a point of principle---the
problematic degeneracies
are
those which occur for
the $\bar{\alpha}^\alpha_m$.
But in practice, for computation reasons,
it is sometimes more practical to use
the $\alpha^\alpha_m$ as a proxy.

To measure the correlation between two primordial
bispectra $B_1$ and $B_2$ we introduce an inner
product, defined by
\begin{equation}
    \ipleft S^1, S^2 \ipright \equiv
    \int_{\mathcal{V}} \d v \;
    S_1(k_1, k_2, k_3)
    S_2(k_1, k_2, k_3)
    \omega(k_1, k_2, k_3) ,
    \label{eq:primordial-inner-product}
\end{equation}
where $S_1$ and $S_2$ are the corresponding shape functions,
$\d v$ is an element of volume on the integration domain $\mathcal{V}$
(which corresponds to allowable triangular configurations of the
momenta $\vect{k}_i$), and
$\omega$ is a weight function which can be chosen to suit our convenience.
For a more detailed discussion of
Eq.~\eqref{eq:primordial-inner-product}
we refer to the literature~\cite{Fergusson:2009nv}.
We normalize the $\Rmode^n$ so that
$\ipleft \Rmode^m, \Rmode^n \ipright = \delta^{mn}$
and therefore~\eqref{eq:shape-decomposition} implies
\begin{equation}
    \ipleft S^1, S^2 \ipright = \sum_{n} \alpha^1_m \alpha^2_m .
\end{equation}
When measuring correlations between angular bispectra
it is helpful to account for the ability of the WMAP instrument
to distinguish between different shapes.
This ability is measured by the matrix $(\Cov^{-1})^{mn}$
discussed in footnote~\ref{footnote:wmap-monte-carlo}
on p.~\pageref{footnote:wmap-monte-carlo}.
We define
\begin{equation}
    \ipleft b^1_{\ell_1, \ell_2, \ell_3} , b^2_{\ell_1, \ell_2, \ell_3} \ipright
    =
    \sum_{m,n} \bar{\alpha}^1_{m} (\Cov^{-1})^{mn} \bar{\alpha}^2_n .
\end{equation}
Note that we write the inner product $\ipleft \cdot , \cdot \ipright$
for both the primordial and angular bispectra,
but they are \emph{not} equal;
the definition is different depending whether it is
taken between primordial or angular bispectra.
In either case it is conventional to measure the correlation between shapes
by defining a `cosine',
\begin{equation}
    \cos (1, 2) =
    \frac{\ipleft 1, 2 \ipright}
    {\ipleft 1 \ipright^{1/2} \ipleft 2 \ipright^{1/2}} ,
\end{equation}
where `1' and `2' should be substituted by the appropriate
angular or primordial bispectrum.

\para{Principal directions}%
To factor out the degenerate directions we diagonalize
$\vect{\Fisher}$,
finding a new orthogonal
matrix $\vect{U}$
and a nonnegative-definite diagonal matrix $\vect{\Sigma}$
so that
$\vect{\Fisher} = \vect{U} \vect{\Sigma} \vect{U}^\transpose$
where a superscript `$\transpose$' denotes matrix transposition.%
    \footnote{In practice, it can happen that numerical inaccuracies
    cause $\vect{\Fisher}$ to develop very small negative eigenvalues which
    spoil simple diagonalization strategies.
    Where this occurs we perform a singular value decomposition,
    which corresponds to finding (possibly complex) unitary
    matrices
    $\vect{U}$, $\vect{V}$
    and a nonnegative-definite diagonal matrix $\vect{\Sigma}$ so that
    $\vect{\Fisher} = \vect{U} \vect{\Sigma} \vect{V}^\transpose$.
    We discard complex directions
    and check that the results are stable under exchange of $\vect{U}$ and
    $\vect{V}$.}
The matrix $\vect{U}$ can be regarded as a rotation from the
operators $\Op^\alpha$ to a new set of operators
$\Op^{\alpha'}$ which satisfy
$\Op^{\alpha'} = \sum_\alpha \Op^\alpha {U_\alpha}^{\alpha'}$,
and likewise a new set of dimensionless coefficients
$\lambda_{\alpha'} = \sum_\alpha \lambda_\alpha {U^\alpha}_{\alpha'}$.
The Lagrangian $\sum_\alpha \lambda_\alpha \Op^\alpha =
\sum_{\alpha'} \lambda_{\alpha'} \Op^{\alpha'}$ is invariant
under a rigid rotation of this kind.

\begin{figure}
        \centering
        \begin{subfigure}[b]{0.4\textwidth}
                \centering
                \includegraphics[width=\textwidth]{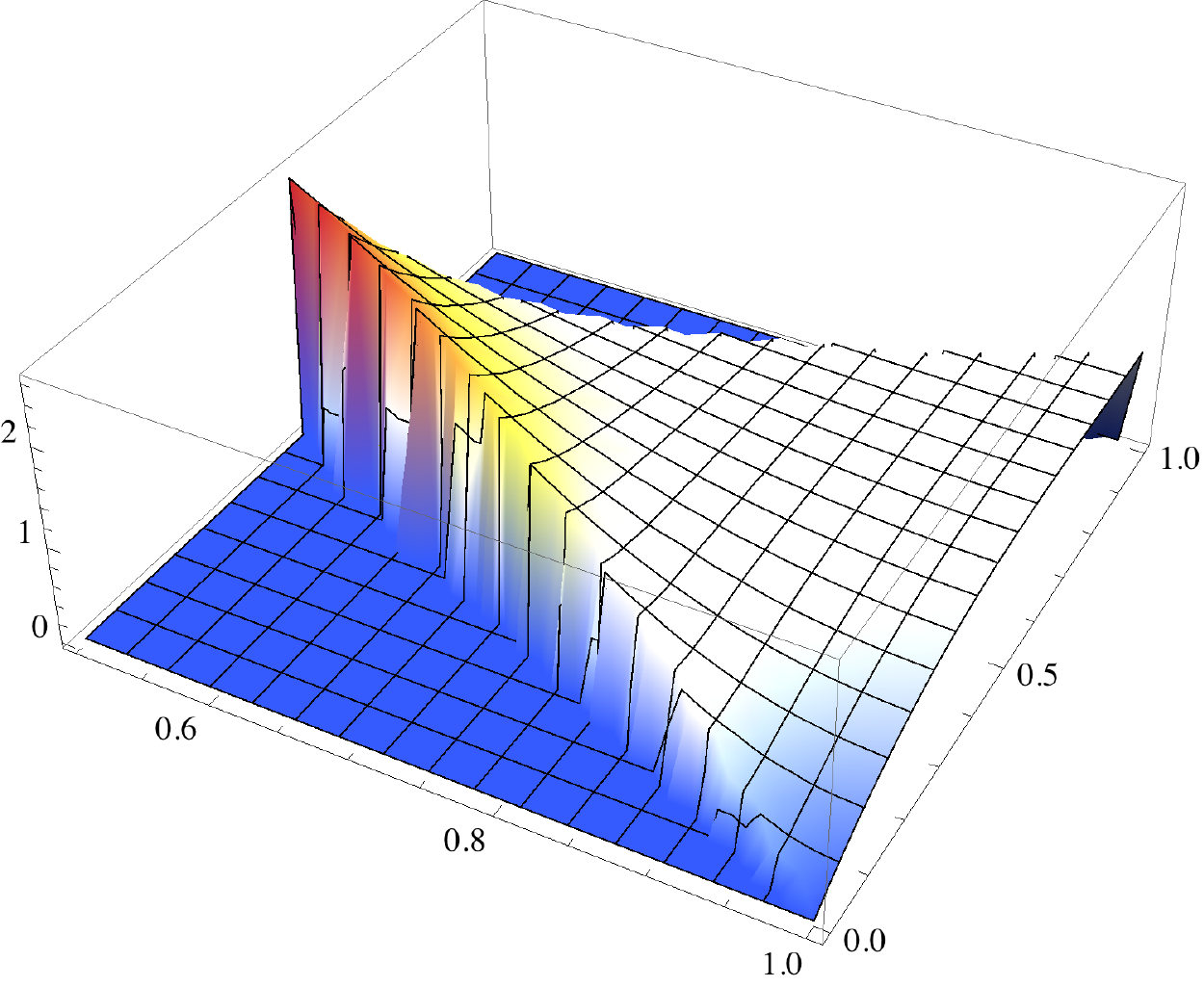}
                \caption{$\Op^1$\label{fig:data-1}}
        \end{subfigure}%
        \qquad \qquad 
        \begin{subfigure}[b]{0.4\textwidth}
                \centering
                \includegraphics[width=\textwidth]{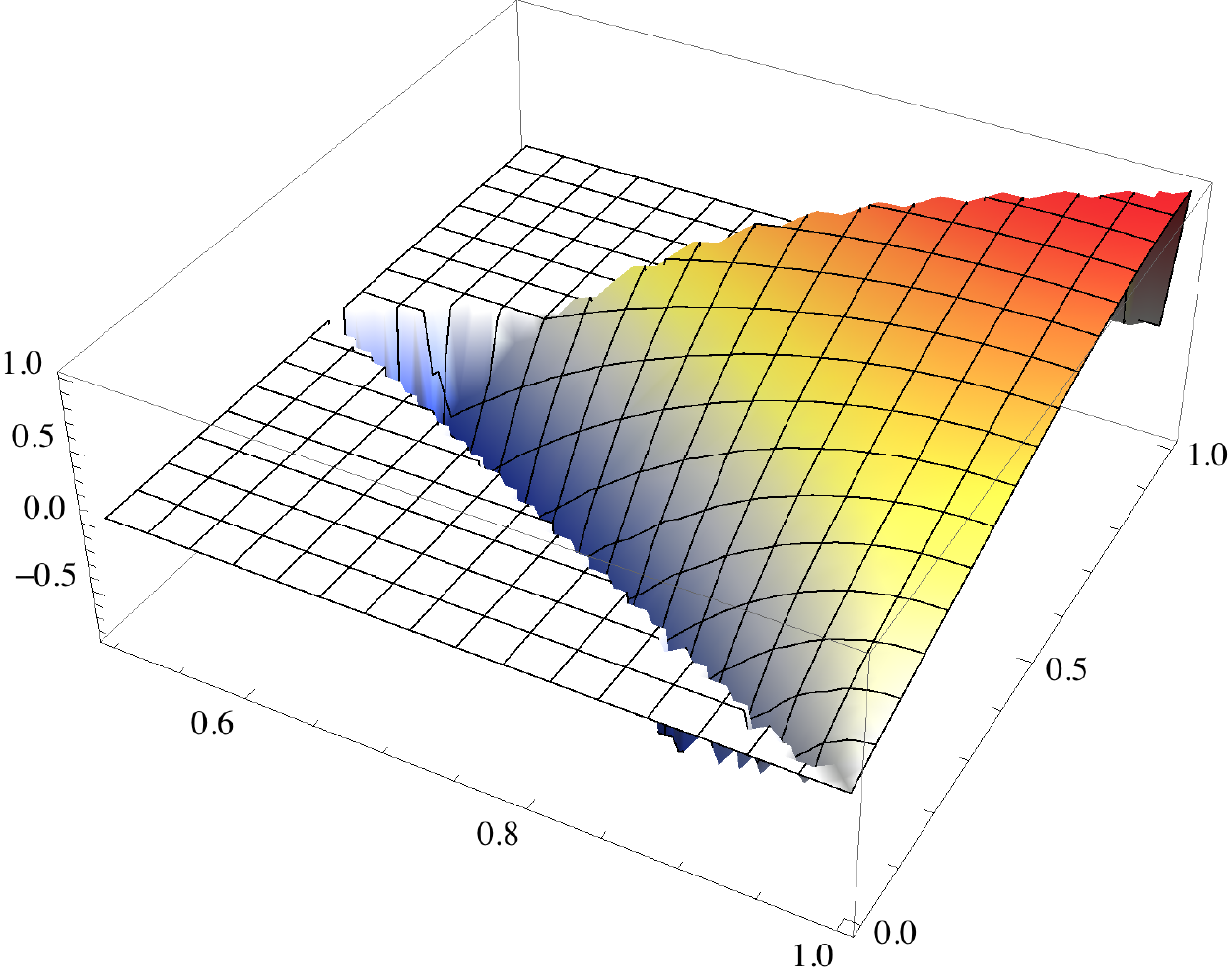}
                \caption{$\Op^2$\label{fig:data-2}}
        \end{subfigure}

        \vspace{5mm} 
        \begin{subfigure}[b]{0.4\textwidth}
                \centering
                \includegraphics[width=\textwidth]{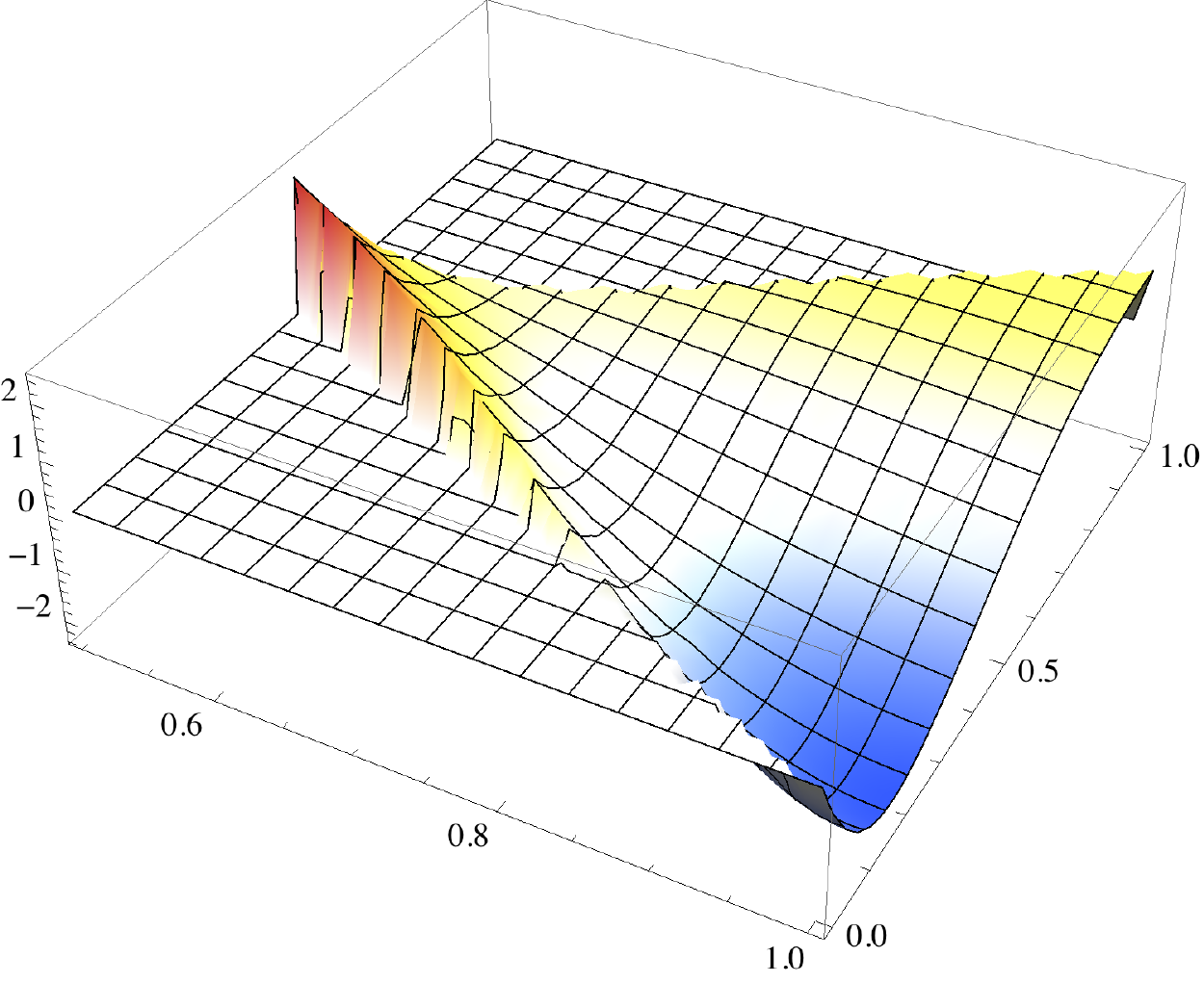}
                \caption{$\Op^3$\label{fig:data-3}}
        \end{subfigure}
        \qquad \qquad \begin{subfigure}[b]{0.4\textwidth}
                \centering
                \includegraphics[width=\textwidth]{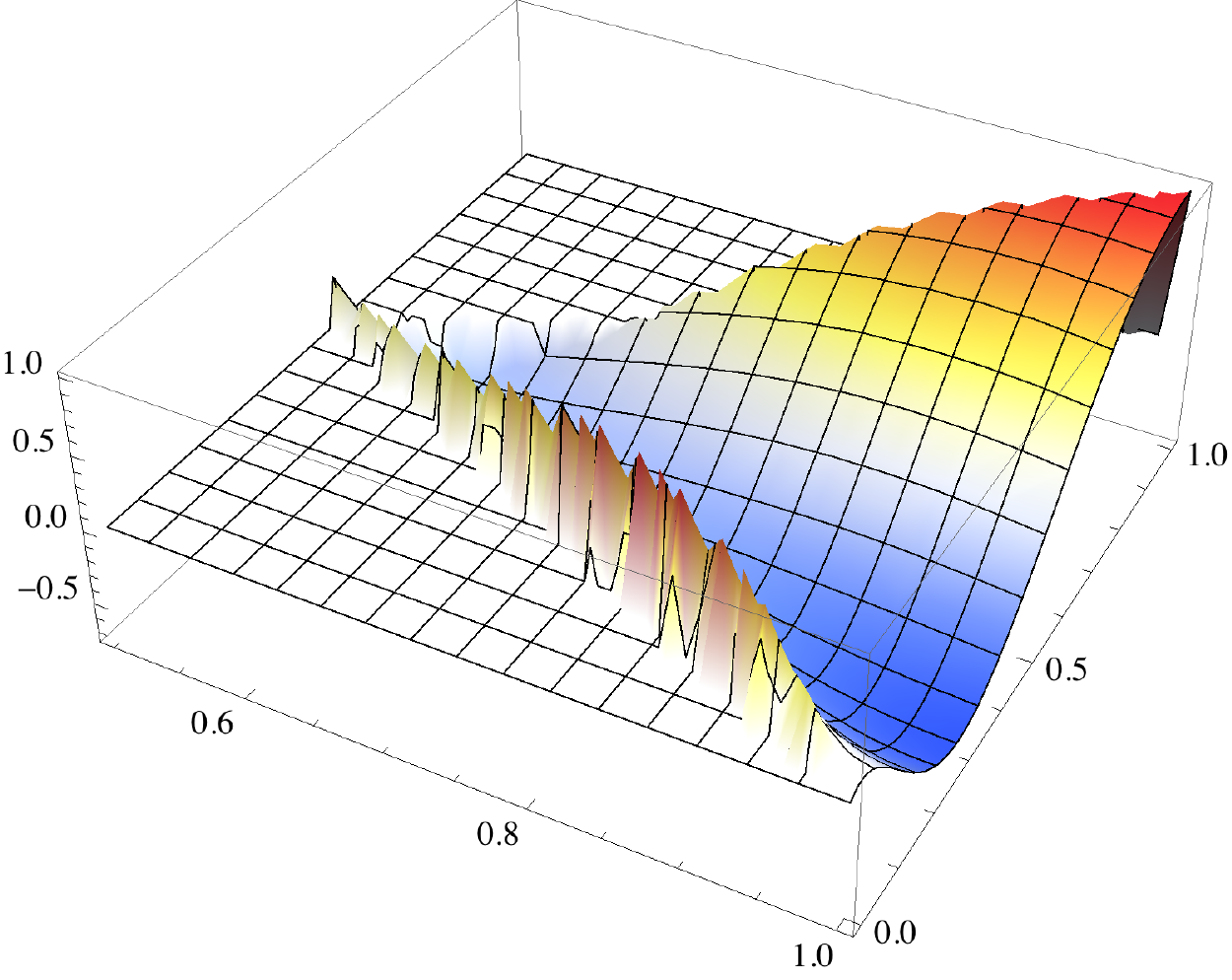}
                \caption{$\Op^4$\label{fig:data-4}}
        \end{subfigure}

        \caption{Bispectrum shapes generated by the operators
        $\Op^{\alpha'}$ corresponding to the constrainable
        parameter combinations $\lambda_{\alpha'}$.
        These plots follow the conventions
        of Babich et al.~\cite{Babich:2004gb}.
        For each
        $\Op^{\alpha'}$ the plotted quantity is $x^2 y^2 B_{\alpha'}(x,y,1) / B_{\alpha'}(1,1,1)$
        where $x = k_1 / k_3$ and $y = k_2 / k_3$
        (no sum on $\alpha'$).
        \label{fig:data}}
\end{figure}

The presence of degeneracies means that the eigenvalues of
$\vect{\Fisher}$ vary significantly in magnitude.
The largest eigenvalues are
\begin{equation}\label{eq:covar}
  2.23 \times 10^{-3}, \,
    1.70 \times 10^{-4}, \,
    7.17 \times 10^{-7}, \,
    1.22 \times 10^{-9}, \; \text{and} \;
    1.20 \times 10^{-14} ,
\end{equation}
with the remaining eigenvalues being of order $10^{-15}$ or smaller.
We retain the first four, which corresponds to a hierarchy between largest
and smallest eigenvalues of $\sim 10^{6}$.
The corresponding eigenvectors in parameter space
yield four linear combinations $\lambda_1$,
$\lambda_2$, $\lambda_3$, $\lambda_4$ which can be constrained.
It should be remembered that because the covariance matrix
$\Cov_{mn}$ defined in~\eqref{eq:covariance-def}
depends on details of the WMAP experiment
(including the masks, beam and noise properties
as described in footnote~\ref{footnote:wmap-monte-carlo}
on p.~\pageref{footnote:wmap-monte-carlo}),
the Fisher matrix and therefore the
shapes corresponding to these leading eigenvalues
also depend on these details.
They may vary between experiments, depending
on the varying sensitivity
of each experiment to different regions of
multipole-space.
The precise specification of the
leading shapes given
in Table~\ref{tab:rotated-lambda},
and we plot the shapes of the corresponding
primordial bispectra in Fig.~\ref{fig:data}.

\para{Shapes of principal directions}%
These shapes can be
given an approximate
interpretation in terms of
the standard templates.
Fig.~\ref{fig:data-1}
is associated with the largest eigenvalue,
and is therefore the best-measured shape.
It
exhibits significant correlations
for $x = y = 0.5$, which corresponds the `folded'
configuration~\cite{Meerburg:2009ys}.
Fig.~\ref{fig:data-2} exhibits significant correlations in the
equilateral limit $x = y = 1$, and some anticorrelation in the
folded configuration.
It can be regarded as an approximate `orthogonal'
shape~\cite{Senatore:2009gt}.
Together, a linear combination of these two configurations
can be used to produce an approximate `equilateral'
shape.
These results are consistent with the forecast of
Byun \& Bean~\cite{Byun:2013jba},
who suggested that (neglecting the local shape),
the highest signal-to-noise should be achieved
for shapes similar to the folded and orthogonal
templates.
Note that
Byun \& Bean's analysis was based on a survey
with Planck-like masks, beams and noise
rather than the WMAP9 characteristics adopted here.

Fig.~\ref{fig:data-3} has an interior node, where---%
without our choice of signs---%
anti-correlations
have a local maximum in the interior of the allowed
triangular region.
This is quite different to the behaviour of
Figs.~\ref{fig:data-1}--\ref{fig:data-2},
in which local maxima only occur for extreme configurations
on the boundary of the allowed region.
The shape of Fig.~\ref{fig:data-3} is similar
to a shape produced in a Galileon theory by
Creminelli et al.~\cite{Creminelli:2010qf},
and later reproduced in a general Horndeski
Lagrangian by Refs.~\cite{Burrage:2011hd,Ribeiro:2011ax}.
Finally, Fig.~\ref{fig:data-4}
is a complex shape containing an interior node
together with substantial correlations in the
equilateral configuration.
It represents something different from the
shape of Fig.~\ref{fig:data-3}, but
it will be seen in~{\S\ref{sec:results}}
below that it is rather weakly constrained by the data.

These results are consistent with the conclusions
of Ribeiro et al.~\cite{Ribeiro:2011ax},
who found that in a very general single-field model%
    \footnote{Ribeiro et al. worked with a model for the fluctuations
    which is equivalent to the fluctuations in a general
    Horndeski action. Although very permissive, this model
    is still less general than the full effective
    field theory~\eqref{EFTLag}.}
it
could be possible to produce a measurable signal in
a mode similar to that of Fig.~\ref{fig:data-3},
or equivalently the Creminelli et al. shape~\cite{Creminelli:2010qf},
but that further orthogonal shapes would be difficult to measure.

\para{Correlation of shapes}%
We tabulate the correlation between these shapes in
Table~\ref{tab:cosines},
and also between these shapes and the standard CMB templates.
The correlation is computed for the primordial bispectra
using~\eqref{eq:primordial-inner-product}.
In particular we note that, although the
angular bispectra for the
$\Op^{\alpha'}$ are orthogonal
by construction,
mapping back to the primordial bispectrum
introduces some correlation;
for example, $\cos( \Op^3, \Op^4) = -0.62$.
This degradation is expected, because
the increasing covariance
represented by~\eqref{eq:covar} will cause noise to
dominate over signal.
Therefore the linear relationship between the 
primordial and CMB $n$-point functions,
implied by Eq.~\eqref{eq:deltaT-alm},
is no longer satisfied.

One can regard these results as a reflection of the fact
that
the first three operators
$\Op^1$, $\Op^2$ and $\Op^3$ are reasonably well-measured,
whereas the fourth operator
$\Op^4$ is only weakly constrained.
\begin{table}
\centering
    \small
	\heavyrulewidth=.08em
	\lightrulewidth=.05em
	\cmidrulewidth=.03em
	\belowrulesep=.65ex
	\belowbottomsep=0pt
	\aboverulesep=.4ex
	\abovetopsep=0pt
	\cmidrulesep=\doublerulesep
	\cmidrulekern=.5em
	\defaultaddspace=.5em
	\renewcommand{\arraystretch}{1.4}

    \rowcolors{2}{gray!25}{white}
	\begin{tabular}{stttt} 
    \toprule
 	 & \Op^1 & \Op^2 & \Op^3 & \Op^4 \tabularnewline
 	\Op^1  & 1.00 & -0.03 & -0.11 & -0.01 \tabularnewline 
	\Op^2 & -0.03 & 1.00 & 0.17 & 0.24 \tabularnewline 
	\Op^3 & -0.11 & 0.17 & 1.00 & -0.62 \tabularnewline 
	\Op^4 & -0.01 & 0.24 & -0.62 & 1.00 \tabularnewline 
	\text{constant} & -0.95 & -0.21 & -0.09 & 0.00\tabularnewline
	\text{equilateral} & -0.80 & -0.57 & 0.03 & -0.16 \tabularnewline
	\text{flat} & -0.73 & 0.19 & -0.38 & 0.31 \tabularnewline
	\text{local} & -0.54 & 0.00 & -0.29 & 0.02 \tabularnewline
	\text{orthogonal} & 0.36 & -0.79 & 0.27 & -0.35 \tabularnewline
	\bottomrule
\end{tabular}
\caption{Cosines of the shapes appearing in Fig.~\ref{fig:data}
between themselves and the standard CMB templates.
Inner products are computed using~\eqref{eq:primordial-inner-product}
and the constant
bispectrum as a reference.\label{tab:cosines}}
\end{table}

\subsection{Results}
\label{sec:results}
A framework for estimating the $\hat{\beta}_n$
from a CMB temperature map
using
wavelet or needlet methods was
developed by Regan et al.~\cite{Regan:2013wwa,Regan:2013jua}.
We apply these methods to 9-year data from the WMAP
satellite~\cite{Hinshaw:2012aka,Bennett:2012zja}.
For the constrainable parameters
$\{ \lambda_{\alpha'} \} = \{ \lambda_1, \lambda_2, \lambda_3, \lambda_4 \}$ we find
\begin{center}
    \small
	\heavyrulewidth=.08em
	\lightrulewidth=.05em
	\cmidrulewidth=.03em
	\belowrulesep=.65ex
	\belowbottomsep=0pt
	\aboverulesep=.4ex
	\abovetopsep=0pt
	\cmidrulesep=\doublerulesep
	\cmidrulekern=.5em
	\defaultaddspace=.5em
	\renewcommand{\arraystretch}{1.5}
        
    \rowcolors{2}{gray!25}{white}

    \begin{tabular}{sd}
        \toprule
        & \multicolumn{1}{c}{Estimate} \\
        \hat{\lambda}_1 & -22.9 ! 20.9 \\
        \hat{\lambda}_2 & 94.9 ! 76.7 \\
        \hat{\lambda}_3 & -956 ! 1180 \\
        \hat{\lambda}_4 & 42400 ! 28600 \\
        \bottomrule

    \end{tabular}
\end{center}
The quoted errors are $1\sigma$ and marginalized over the other
$\lambda_{\alpha'}$.
\begin{sidewaystable}
\centering
    \small
	\heavyrulewidth=.08em
	\lightrulewidth=.05em
	\cmidrulewidth=.03em
	\belowrulesep=.65ex
	\belowbottomsep=0pt
	\aboverulesep=.4ex
	\abovetopsep=0pt
	\cmidrulesep=\doublerulesep
	\cmidrulekern=.5em
	\defaultaddspace=.5em
	\renewcommand{\arraystretch}{2.4}

    \rowcolors{2}{gray!25}{white}

    \begin{tabular}{suuuuuuuuuuu}
        \toprule
        & \multicolumn{1}{t}{\lambda_A}
        & \multicolumn{1}{t}{\lambda_B}
        & \multicolumn{1}{t}{\lambda_C}
        & \multicolumn{1}{t}{\lambda_D}
        & \multicolumn{1}{t}{\lambda_E}
        & \multicolumn{1}{t}{\lambda_F}
        & \multicolumn{1}{t}{\lambda_G}
        & \multicolumn{1}{t}{\lambda_H}
        & \multicolumn{1}{t}{\lambda_I}
        & \multicolumn{1}{t}{\lambda_J}
        & \multicolumn{1}{t}{\lambda_K} \\

        \lambda_1 & -0.173206 & -0.193711 & -0.221619 & -0.260354 & -0.203464 & -0.260354
            & -0.256692 & -0.507151 & -0.24732 & -0.446539 & 0.350338 \\

        \lambda_2 & -0.240889 & -0.222667 & -0.200172 & -0.163445 & -0.215441 & -0.163445
            & -0.165617 & 0.060200 & -0.171386 & 0.048317 & -0.830493 \\
            
        \lambda_3 & 0.269654 & 0.191121 & 0.050094 & -0.064110 & 0.132429 & -0.064110
            & 0.006499 & -0.782966 & 0.226131 & 0.379275 & -0.233301 \\
            
        \lambda_4 & -0.219333 & -0.233381 & -0.193386 & -0.279000 & -0.203131 & -0.279000
            & -0.120479 & 0.176132 & 0.481877 & 0.541142 & 0.304148 \\

        \bottomrule
    \end{tabular}

    \caption{Linear combinations of the $\lambda_\alpha$ parameters
    which can be constrained using the WMAP9 bispectrum data.
    The unrotated parameters are labelled $\lambda_A$, $\lambda_B$, \ldots,
    and correspond to those
    defined in
    Table~\ref{tab:coeff}
    in terms of the EFT mass scales.
    The rotated parameters are labelled $\lambda_1$, $\lambda_2$, \ldots.
    \label{tab:rotated-lambda}}
\end{sidewaystable}

Senatore, Smith \& Zaldarriaga~\cite{Senatore:2009gt}
obtained constraints on the amplitude of the `equilateral' and `orthogonal'
bispectrum templates from the 5-year WMAP data,
and used these to constrain a subset of terms in the effective
Lagrangian~\eqref{EFTLag}.
They concluded that
each shape included in their analysis
could be approximately described by a linear combination of these
two templates, up to $\sim 90\%$ correlation.
However, they
included only two of the operators in Eq.~\eqref{EFTLag}.
Our analysis demonstrates that it is possible to increase the number
of linearly independent operators from two to four,
although the estimate $\hat{\lambda}_4 = 42400 \pm 28600$ shows that
that sensitivity is already decreasing markedly for the fourth parameter.

\section{Constraints on models}
\label{sec:model-constraints}
In~{\S\ref{sec:estimating}} we obtained constraints on certain linear
combinations of the EFT scales $M_i$, $\bar{M}_i$.
The remaining linear combinations formally have
infinite uncertainties because of degeneracies.
Together, these results summarize the information which can
be recovered from the WMAP9 bispectrum, but to apply them to specific
models we must first match the mass scales $M_i$, $\bar{M}_i$.
In this section we give two examples of this programme for
models of observational interest:
the Dirac--Born--Infeld model (`DBI inflation')
and `Ghost inflation'.

Once the $M_i$, $\bar{M}_i$ are known,
the results of~{\S\ref{sec:estimating}} would give
four constraints on different combinations of these scales.
Depending how many scales are needed to parametrize
an individual model, it may be possible to estimate
some or all of them, or they may even be over-constrained.
The latter possibility indicates that the model
is a poor fit for the data.
Where more than four mass scales are needed to characterize
a model, the constraints pick out
an observationally-allowed subspace which is consistent with
the CMB bispectrum measurements.

\para{Methodology}
To map our four constraints
for $\{ \lambda_1, \lambda_2, \lambda_3, \lambda_4 \}$
onto a subset of the original parameter space
$M_i$, $\bar{M}_i$
we minimize the value
\begin{equation}
    X^2 = \sum_{\alpha'=1}^4
    \frac{\big[\lambda_{\alpha'}(M_i, \bar{M}_i) - \hat{\lambda}_{\alpha'}\big]^2}
    {\Sigma_{\alpha'\alpha'}^{-1}} ,
    \label{eq:chi-square}
\end{equation}
where $\Sigma_{\alpha'\beta'}$ is the diagonal
matrix of principal eigenvalues
listed in Eq.~\eqref{eq:covar};
$\lambda_{\alpha'}(M_i, \bar{M}_i)$ represents
the value of the linear combination $\lambda_{\alpha'}$ which
would be predicted given a fixed choice of mass scales
$M_i$, $\bar{M}_i$;
and $\hat{\lambda}_{\alpha'}$ represents the
value estimated from the data in~{\S\ref{sec:estimating}}. 
With four constraints on the $\lambda_{\alpha'}$ we can constrain up to
four of the $M_i$, $\bar{M}_i$.

We take $X^2$ to be $\chi^2$-distributed with four degrees of freedom.
The $\lambda_{\alpha'}$ are constructed from
a linear combination
of the $\hat{\beta}_n$, and we assume that the experimental
error for each $\hat{\beta}_n$ is independent and Gaussian-distributed.
Because the $\lambda_{\alpha'}$ are chosen to be orthogonal,
the experimental errors on each $\hat{\lambda}_{\alpha'}$
will be obtained from a nearly uncorrelated sum of Gaussians,
and will therefore also be nearly independent.
This makes $X^2$ approximately equal to a sum of four
approximately independent,
unit Gaussians,
and hence roughly $\chi^2$-distributed.

Confidence intervals for the $M_i$, $\bar{M}_i$
could be determined by searching
for suitable critical values of the $\chi^2$ distribution.
Alternatively, assuming that the $\hat{\lambda}_{\alpha'}$
have uncorrelated Gaussian errors,
we could expand $X^2$ to second order around the maximum likelihood point,
\begin{equation}
\begin{split}
    X^2 & = X^2\big|_{\text{\textsc{mle}}}
    + \sum_{\alpha'\beta'} \left.\frac{\partial^2 X^2}
    {\partial \lambda_{\alpha'} \partial \lambda_{\beta'}}
    \right|_{\text{\textsc{mle}}}
    \big(\lambda_{\alpha'} - \lambda_{\alpha'}\big|_{\text{\textsc{mle}}}\big)
    \big(\lambda_{\beta'} - \lambda_{\beta'}\big|_{\text{\textsc{mle}}}\big)
    + \cdots
    \\ & = X^2\big|_{\text{\textsc{mle}}} + \Delta X^2 ,
\end{split}
\label{eq:second-order-chisq}
\end{equation}
and construct confidence contours
at the $n^{\text{th}}$-$\sigma$ level
by searching for critical values
where
$\Delta X^2 = n^2$.
In principle these methods agree if the
$\hat{\lambda}_{\alpha'}$ are Gaussian and uncorrelated.
We find that the agreement is not quite exact, which we ascribe
to a small residual correlation between the errors on the
$\hat{\lambda}_{\alpha'}$.
The single-parameter constraints reported below are
obtained using the second-order expansion~\eqref{eq:second-order-chisq},
which reproduces the Fisher-matrix estimates.
For two or more parameters we
report constraints extracted from critical values of the full
$\chi^2$-distribution with $4$ degrees of freedom.

\para{DBI inflation}%
The first example we consider is the `Dirac--Born--Infeld'
or `DBI' inflationary model.

The Dirac--Born--Infeld action describes
fluctuations of a membrane moving in a warped
transverse space, or `throat'. Under certain circumstances
it can describe an inflationary epoch in which
inflaton perturbations propagate at less than the speed
of light from the perspective of a brane-based
observer, due to constraints imposed by
the extradimensional covering theory.
The small sound speed means that
these models can produce significant nongaussianities
in the equilateral mode.

Fluctuations in a single-field DBI model
can be described by the effective
action~\eqref{EFTLag},
retaining only the $B$ and $D$ operators,
\begin{subequations}
\begin{align}
    \label{eq:dbi-op-1}
    \lambda_B \Op^B & \propto M_2^4 \frac{1}{a^2} \dot{\pi} ( \partial \pi )^2 \\
    \label{eq:dbi-op-2}
    \lambda_D \Op^D & \propto M_3^4 \dot{\pi}^3 .
\end{align}
\end{subequations}
This model does not involve the problematic
scales $\bar{M}_2$, $\bar{M}_3$ which lead to normalization
inaccuracies for the single-particle mode functions
and therefore we expect our estimates to be quantitatively
reliable.

The original DBI model had a single free parameter and therefore
$M_2$ and $M_3$ cannot be chosen independently but are correlated
as described below.
Alternatively, one can consider a larger family of DBI-like models
which retain only these operators but allow
$M_2$ and $M_3$ to vary.
Following Senatore et al., constraints are typically
expressed using the parameters
\begin{subequations}
\begin{align}
    \frac{1}{\cs^2} & = 1 - \frac{2 M_2^4}{\Mp^2 \dot{H}}
        = 1 - \frac{324}{85} \lambda_B , \\
    \cthree \Big( \frac{1}{\cs^2} - 1 \Big) & =
        \frac{2 M_3^4 \cs^2}{\Mp^2 \dot{H}}
        = - \frac{243}{10} \lambda_D .
\end{align}
\end{subequations}
Causality requires the speed of sound $\cs$ to be less than
unity.
Since $\dot{H}$ < 0 during inflation it follows
that $M_2^4$ must be positive
(see footnote~\ref{footnote:mass-signs} on p.~\pageref{footnote:mass-signs}).
If $M_2^4 \gtrsim \Mp^4 |\dot{H}|$
then a significant bispectrum can be generated.
The reason for expressing constraints in
terms of these parameters is that it is not possible
to determine $M_2$ and $M_3$ without simultaneously
specifying $\dot{H}$.
The original DBI model imposes
the constraint $\cthree = 3 (1-\cs^2)/2$.

We estimate the \emph{joint} constraints on $\lambda_B$ and $\lambda_D$ to be
\begin{subequations}
\begin{align}
    \label{eq:dbi-est-1}
    \lambda_B & = -1151 \pm 760 \\
    \label{eq:dbi-est-2}
    \lambda_D & =\quad\,\, 946 \pm 584 .
\end{align}
\end{subequations}
The Planck collaboration expressed their constraints in terms
of $\fNL$-like parameters
$\fNLEFTone$ and $\fNLEFTtwo$.
In our notation these correspond, respectively,
to $\lambda_B$ under the assumption
$\lambda_D = 0$
and $\lambda_D$ under the assumption
$\lambda_B = 0$.
Using the 2013 dataset, the Planck collaboration reported the bounds
$\fNLEFTone = 8 \pm 73$ and $\fNLEFTtwo = 19 \pm 57$~\cite{Ade:2013ydc}.
Using the same notation, we find
\begin{subequations}
\begin{align}
    \fNLEFTone & = 68.3 \pm 103 \\
    \fNLEFTtwo & = 69.3 \pm 79 .
\end{align}
\end{subequations}
The Planck2013 errors represent an improvement of order $30\%$.

\begin{figure}
  \caption{Constraints on the DBI-like parameters $\cs$, $\cthree$.
  \label{fig:dbi-like}}
  \centering
  \includegraphics[width=0.7\textwidth]{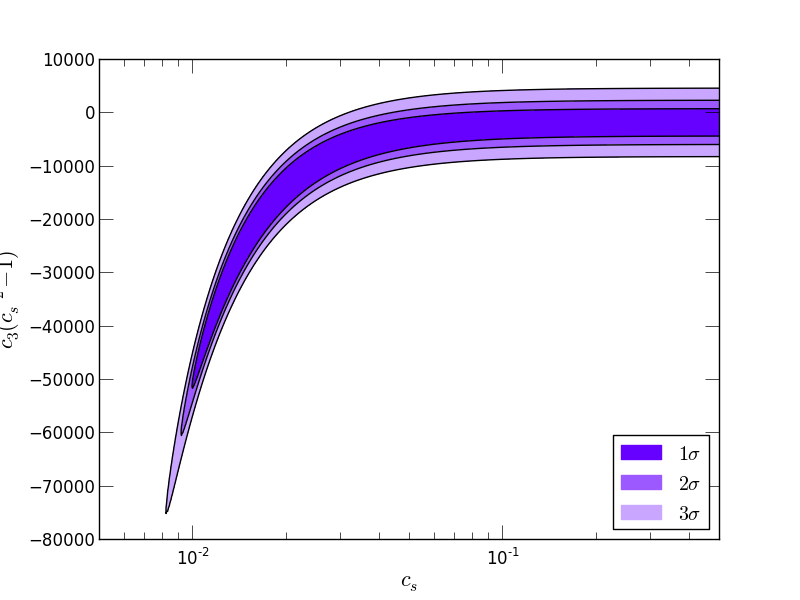}
\end{figure}

Alternatively, each bound can be expressed in terms of
$\cs$ and $\cthree$.
To compare with the constraints
reported by the Planck collaboration
we consider three possibilities.
First, marginalizing over $\cthree$
gives a conservative lower bound on $\cs$,
\begin{subequations}
\begin{equation}
    \cs \geq 0.010 \quad \text{at $95\%$ confidence.}
\end{equation}
For comparison, Planck2013 found $\cs \geq 0.02$~\cite{Ade:2013ydc}
at the same confidence level.
Second, imposing $\cthree = 0$ gives
\begin{equation}
    \cs \geq 0.044 \quad \text{at $95\%$ confidence.}
\end{equation}
Finally, assuming the strict DBI relation between
$\cs$ and $\cthree$ leaves $\cs$ as a single
free parameter. We find
\begin{equation}
    \cs \geq 0.051 \quad \text{at $95\%$ confidence.}
    \label{eq:dbi-cs-constraint}
\end{equation}
\end{subequations}
Planck2013 obtained $\cs \geq 0.07$~\cite{Ade:2013ydc},
also at $95\%$ confidence.
Eq.~\eqref{eq:dbi-cs-constraint}
can
also be expressed as an
$\fNL$ parameter for the DBI shape.
This gives
\begin{equation}
    \fNLDBI = 69.6 \pm 97.4 .
\end{equation}
Finally,
allowing both $\cs$ and $\cthree$ to vary
results in
a lower bound for $\cs$ and
relatively weak constraints for $\cthree$,
plotted in Fig.~\ref{fig:dbi-like}.

Similar bounds were reported by Senatore et al.~\cite{Senatore:2009gt}.
Our construction 
ensures that the bounds reported above
correspond to the most accurate constraints
which can be achieved using this data set,
because the shapes are explored
using four rather than two orthogonal directions
in the likelihood~\eqref{eq:chi-square}.
For example, using only the leading principal component to construct
the likelihood yields the constraint
$\cs >0.061$ in the DBI model.
This bound unduly weights the component of the DBI shape along this principal direction,
giving an overly optimistic constraint when compared
with the four-component result~\eqref{eq:dbi-cs-constraint}.

\para{Ghost inflation}%
Our second example is the `Ghost inflation' model
proposed by Arkani-Hamed et al.~\cite{ArkaniHamed:2003uz},
in which inflation is driven by
a so-called `ghost condensate'
which spontaneously breaks
Lorentz invariance in
the background.
The ghost condensate is described by a scalar field
$\phi$ whose time derivative gains a nonvanishing
vacuum expectation value, $\langle \dot{\phi}^2 \rangle = M^2 \neq 0$.
This expectation value is time-independent
and is not
diluted as inflation proceeds.

In the effective theory,
fluctuations around the background
correspond to nonzero $\bar{M}_2^2$ and
$\bar{M}_3^2$, and the limit $\dot{H} \rightarrow 0$.
This limit sets the quadratic spatial-derivative terms
in~\eqref{EFTLag} to zero, so that the
speed of sound is formally zero.
The fluctuations are nevertheless propagating modes because
higher-order spatial derivative terms are present in the
Lagrangian.
The relevant EFT operators are
\begin{subequations}
\begin{align}
    \label{eq:ghost-op-1}
    \lambda_C \Op^C & \propto \frac{\bar{M}_2^4}{a^5}
        \Big(
            \frac{H}{2} \partial^2 \pi (\partial \pi)^2
            +
            \dot{\pi} \partial^2 \partial_i \pi \partial_i \pi
        \Big) \\
    \label{eq:ghost-op-2}
    \lambda_E \Op^E & \propto \frac{\bar{M}_3^2}{a^4}
        \Big(
            H \partial^2 \pi (\partial \pi)^2
            +
            \dot{\pi} \partial^2 \partial_i \pi \partial_i \pi
        \Big) .
\end{align}
\end{subequations}
The arrangement of derivatives is identical up to a relative factor of
2 in the first term.  In terms of the mass scales
$\bar{M}_2$ and $\bar{M}_3$
we have
\begin{subequations}
\begin{align}
    \label{eq:ghost-C}
    \lambda_C & = -\frac{325}{972 \cs^2\epsilon} \frac{\bar{M}_2^2}{\Mp^2} , \\
    \label{eq:ghost-E}
    \lambda_E & = -\frac{130}{243 \cs^2\epsilon} \frac{\bar{M}_3^2}{\Mp^2} .
\end{align}
\end{subequations}
The inclusion of factors of $\cs$ and $\epsilon$ is purely formal,
since this model technically involves the limits $\cs \rightarrow 0$
and $\epsilon \rightarrow 0$.
Proceeding as for the DBI model we obtain the maximum-likelihood
estimates
\begin{subequations}
\begin{align}
    \label{eq:ghost-est-1}
    \lambda_C & = -3680 \pm 2280 \\
    \label{eq:ghost-est-2}
    \lambda_E & = 3900 \pm 2450 .
\end{align}
\end{subequations}

Bartolo et al. observed that
the operators~\eqref{eq:ghost-C}--\eqref{eq:ghost-E}
are nearly the same
and chose to aggregate them into a single
term operator by
introducing a common mass scale $\bar{M}_0$,
satisfying by
$\bar{M}_0^2\equiv 2 \bar{M}_3^2/3 \equiv -2 \bar{M}_2^2$.
The $\lambda$ corresponding to this aggregate
operator [still defined to satisfy the
normalization condition~\eqref{eq:bispectrum-normalization}]
represents an estimate of the amplitude of the
ghost-inflation bispectrum,
and we label it $\lambdaghost$.
As explained in~{\S\ref{sec:bispec}},
the ghost inflation model involves fourth-order
kinetic terms whose details we do not capture,
and therefore the precise normalization
of this estimate is uncertain. We find
\begin{equation}
    \lambdaghost = -68.4 \pm 100.5 .
\end{equation}
For comparison, the Planck collaboration reported
the constraint
$\fNLghost = -23 \pm 88$.
Both estimates agree that the bispectrum in this channel
is consistent with zero within $1\sigma$.
In addition,
this comparison shows that, even in a case where
the normalization uncertainty is important,
our result matches
an exact calculation within a factor of order unity.

\begin{figure}
\centering
\begin{subfigure}{.45\textwidth}
  \includegraphics[width=.75\linewidth]{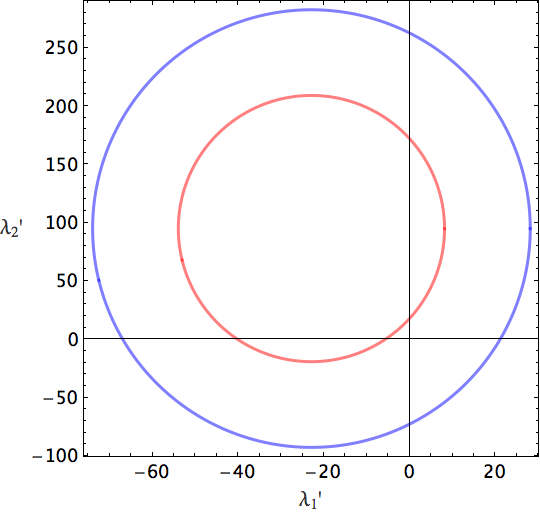}
  \caption{$\lambda_{1}^{\prime}$ versus $\lambda_{2}^{\prime}$}
  \label{fig:sub1}
\end{subfigure}%
\centering
\begin{subfigure}{.45\textwidth}
  \includegraphics[width=.75\linewidth]{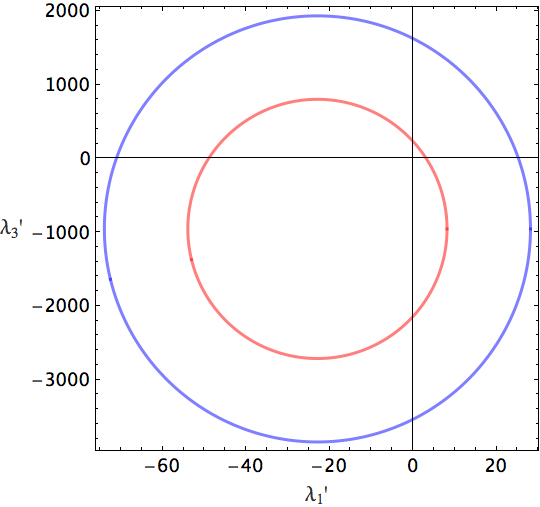}
  \caption{$\lambda_{1}^{\prime}$ versus $\lambda_{3}^{\prime}$}
  \label{fig:sub2}
\end{subfigure}
\centering
\begin{subfigure}{.45\textwidth}
  \includegraphics[width=.75\linewidth]{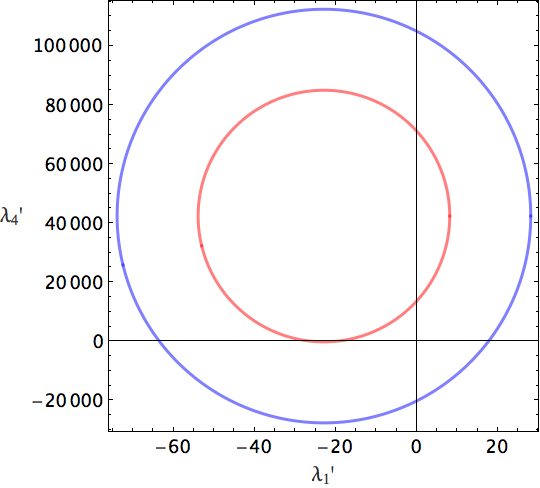}
  \caption{$\lambda_{1}^{\prime}$ versus $\lambda_{4}^{\prime}$}
  \label{fig:sub3}
\end{subfigure}
\centering
\begin{subfigure}{.45\textwidth}
  \includegraphics[width=.75\linewidth]{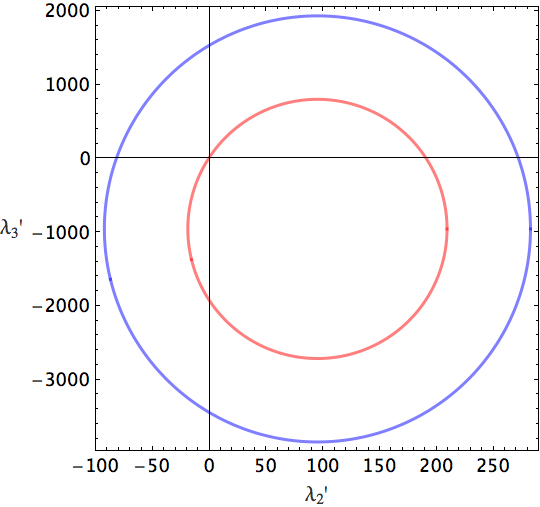}
  \caption{$\lambda_{2}^{\prime}$ versus $\lambda_{3}^{\prime}$}
  \label{fig:sub4}
\end{subfigure}
\centering
\begin{subfigure}{.45\textwidth}
  \includegraphics[width=.75\linewidth]{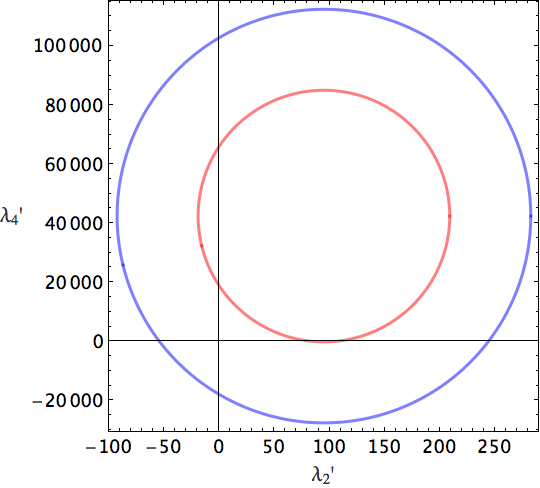}
  \caption{$\lambda_{2}^{\prime}$ versus $\lambda_{4}^{\prime}$}
  \label{fig:sub5}
\end{subfigure}
  \centering
\begin{subfigure}{.45\textwidth}
  \includegraphics[width=.75\linewidth]{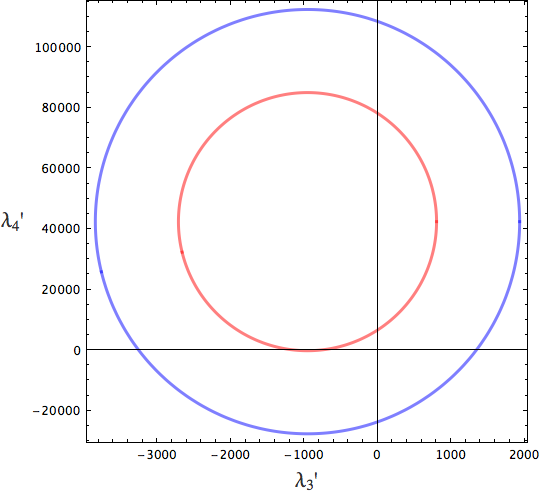}
  \caption{$\lambda_{3}^{\prime}$ versus $\lambda_{4}^{\prime}$}
  \label{fig:sub6}
\end{subfigure}
\caption{$1\sigma$ (red) $2\sigma$ (blue) confidence regions for two out of four principal components $\hat{\lambda}_{\alpha^{\prime}}$ constrained by 9-year WMAP data. The results show consistency with zero magnitude generally within $1-1.5 \sigma$, suggesting no strong evidence of nongaussianity in the single field inflationary parameter space.}
\label{fig:test}
\end{figure}

\section{Model comparison using the bispectrum}
\label{sec:model-comparison}

The analyses of~{\S\S\ref{sec:estimating}--\ref{sec:model-constraints}}
determine best-fit values for the $\lambda_\alpha$,
essentially in a frequentist sense,
assuming a fixed
model for the underlying microphysical fluctuations.
Therefore our conclusions up to this point
are restricted to \emph{parameter estimation}.

Within this framework it is not possible to address questions
such as whether the best-fit combination for
DBI inflation, Eqs.~\eqref{eq:dbi-est-1}--\eqref{eq:dbi-est-2},
represents a better description of the data than the
best-fit combination for Ghost inflation, Eqs.~\eqref{eq:ghost-est-1}--\eqref{eq:ghost-est-2}.
These broader questions constitute the province
of \emph{model comparison}.
Recent work has addressed the issue of
inflationary
model comparison based on
measurements of the two-point function
of the temperature anisotropy~\cite{Martin:2013nzq}.
It is much more challenging to perform a similar
analysis based on the three-point function.
In this section we take some steps towards this objective
within the framework described
in~{\S\S\ref{sec:EFT}--\ref{sec:estimating}}.

For computational
reasons we must impose limitations on the meaning
of the term `model'.
Conceptually this should include whatever information
is necessary to specify the value of each observable.
For example, the transfer matrix ${\Gamma_n}^m$ depends on the
post-inflationary cosmological history
and
in a global analysis the parameters which specify this
history should be varied in addition to the inflationary
parameters $M_i$, $\bar{M}_i$.
However, this generates a large parameter space
which is expensive to search
because
calculation of ${\Gamma_n}^m$
is time-consuming.
In this analysis we will fix the transfer matrix
using standard best-fit values for the
post-inflationary history
and address the
more restricted
question of which inflationary model
yields a better fit for the three-point function
given these assumptions.

\para{Evidence for a model}%
There is no single metric
which unambiguously quantifies the
evidence
for or against a particular model.
One choice is the `Bayes factor'.
For a particular set of observations $D$
and a pair of models $M_1$ and $M_2$, this is
defined to be the ratio of likelihoods,
\begin{equation}
    \label{eq:bayes-factor}
    K_{12} = \frac{P(D | M_1)}{P(D | M_2)}
    =
    \frac{\int P(D | \lambda_1, M_1) P(\lambda_1 | M_1) \, \d \lambda_1}
    {\int P(D | \lambda_2, M_2) P(\lambda_2 | M_2) \, \d \lambda_2} .
\end{equation}
We use $\lambda_1$, $\lambda_2$ to schematically
denote two different choices of the
parameters $\lambda_\alpha$
which characterize a particular model.
Because $M_1$ and $M_2$ are different
they may require a different number of parameters.

The probabilities $P(\lambda_i | M_i)$
represent the prior probability, for each
model, that a particular
parameter choice occurs.
They
must be chosen
by hand.
Where meaningful prior information exists (for example,
previous measurements of a parameter)
this can be encoded using these probabilities.
But
their arbitrariness implies that---%
unless it happens that $K$
is nearly independent of the
$P(\lambda | M)$---%
the Bayes factor is not easy to interpret.
Usually, $K$ is independent of the priors only
when the data are very constraining.
In what follows
we will see that the 9-year WMAP bispectrum data
are insufficiently constraining for this to occur,
so that ambiguities in the interpretation of $K$ remain.

Empirical scales
are used to give meaning to the Bayes factor.
Commonly used examples are due to Jeffreys
or Kass \& Raftery~\cite{doi:10.1080/01621459.1995.10476572}.
In Kass \& Raftery's prescription,
$\ln K$ in the range $(1,3)$ is considered
evidence in favour of $M_1$,
whereas
$\ln K$ in the range $(3,5)$ is considered strong
evidence
and
larger values of $K$ are considered decisive.
Ratios
for which $|\ln K| < 1$ are uninformative.

\para{Choice of priors}%
In our case the $\lambda_\alpha$ represent
Lagrangian coefficients. Some prior estimates exist,
but the datasets from which these were obtained are
not independent of the 9-year WMAP data used in this
analysis. For this reason we disregard these prior constraints,
and therefore
some other way must be found to
justify the functional form of each prior.

If we insist that perturbation theory is valid
then the $\lambda_\alpha$ should not be too large.
This requirement is convenient but not
obviously necessary.
However,
for the purpose of performing a concrete calculation
we shall adopt it in what follows.
In that case, the requirement that the bispectrum generated
by the operator $\Op^\alpha$ does not overwhelm the
power spectrum
$\Pzeta$
is roughly $|\lambda_\alpha| \Pzeta^{1/2} \lesssim 1$,
and therefore
$|\lambda_\alpha| \lesssim 10^4$.
This limit is helpful but
gives no guidance regarding the functional form
of $P(\lambda_\alpha | M)$.
To explore the range of outcomes we consider
two possibilities:
\begin{itemize}
    \item The `Jeffries prior' $P(\lambda_\alpha) \propto |\lambda_\alpha|^{-1}$.
    This choice assigns equal probability for each decade of
    $|\lambda_\alpha|$: that is, for $\lambda_\alpha$ to be between
    $1$ and $10$, $10$ and $100$, and so on.
    The Jeffries prior makes it relatively likely for $|\lambda_\alpha|$ to be
    near zero, and therefore can be regarded as conservative.%
        \footnote{Strictly, the Jeffries prior has a divergence at
        $\lambda_\alpha = 0$.
        We regularize this by cutting out the region
        $|\lambda_\alpha| < 1$ and taking $P(\lambda_\alpha)$
        to be zero within it. We have checked that our
        results are robust to modest changes of the boundary value. The choice of cutoff at unity is, of course, somewhat arbitrary. However, motivated by the fact that in the single parameter case $\lambda_\alpha$ corresponds to the conventionally defined $\fNL$ parameter, we note that error bars on $\fNL$ for single field inflationary models are at best expected to achieve values of order unity. Therefore, we shall regard $\lambda_\alpha=1$ as a `natural' cutoff, but shall also consider the dependency of the results on the cutoff, by presenting results with cutoff at $0.01$, i.e. at a value of order of the slow roll parameters.   \label{footnote:jeffries}}
    
    \item The flat prior, for which $P(\lambda_\alpha)$ is constant.
    This choice assigns equal probability to each value of $\lambda_\alpha$,
    and therefore makes it relatively more likely for $|\lambda_\alpha|$ to be large.
    It is less conservative than the Jeffries
    prior in the sense that
    that it enhances the probability for the Lagrangian~\eqref{EFTLag}
    to predict observably large nongaussianities.
\end{itemize}

\paragraph{Examples}
\begin{itemize}
    \item First, consider the comparison between a trivial
    Gaussian model ($M_1$) for which $\lambda_\alpha = 0$
    and a DBI-like model ($M_2$) with free parameter $\lambda_B \neq 0$.

    Adopting the Jeffries prior,
    the Bayes factor between these models is
    \begin{equation}
        \label{eq:bayes-dbi}
        K_{21} =
            \frac{1}{I}
            \exp\Big(
                \frac{\chi^2(\lambda_B = 0)}{2}
            \Big)
            \int_{-10^4}^{10^4}
            \exp\Big(
                {-\frac{\chi^2(\lambda_B)}{2}}
            \Big)
            \frac{\d \lambda_B}{|\lambda_B|} .        
    \end{equation}
    The normalization factor $I$ formally satisfies
    $I = \int_{-10^4}^{10^4} \d \lambda_B / |\lambda_B|$,
    although it is regularized as described in
    footnote~\ref{footnote:jeffries}.
    We find $\ln K_{21} \approx 0.63$ which gives no
    preference for either model.
    Note that there is an `Ockham's razor' penalty
    implicit in~\eqref{eq:bayes-dbi},
    because the parameter $\lambda_B$ is allowed to float
    over a relatively large interval.
    For a flat prior this Ockham penalty strongly disfavours the
    DBI model, producing $\ln K_{21} = -4.23$.
    We conclude that the data are not sufficient to overcome
    the ambiguity in specifying a prior.
    
    The Bayes factor $K_{21}$ is only one of a number
    of metrics which can be used to assess goodness of fit.
    Another is the Akaike `information criterion',
    defined by $\AIC = \chisqmle + 2k$, where
    $k$ measures the number of parameters in the model
    and is a proxy for the `Ockham razor' penalty
    of Eq.~\eqref{eq:bayes-factor}.
    The model with smallest $\AIC$ is preferred.
    We find $\AIC_1 - \AIC_2
    = -1.56$, which implies a preference for the trivial
    Gaussian model $M_1$ in comparison to a model with nonzero
    $\lambda_B$.
    The same preference is found if we allow any other
    single Lagrangian parameter to be nonzero.
    
    \item Next, consider a third DBI-like model $M_3$
    in which the two parameters $\lambda_B$
    and $\lambda_D$
    (or, equivalently, the parameters
    $\cs$ and $\cthree$ in the notation of~{\S\ref{sec:model-constraints}})
    are allowed to float.
    We find
    \begin{center}
        \small
    	\heavyrulewidth=.08em
    	\lightrulewidth=.05em
    	\cmidrulewidth=.03em
    	\belowrulesep=.65ex
    	\belowbottomsep=0pt
    	\aboverulesep=.4ex
    	\abovetopsep=0pt
    	\cmidrulesep=\doublerulesep
    	\cmidrulekern=.5em
    	\defaultaddspace=.5em
    	\renewcommand{\arraystretch}{1.3}
    
        \rowcolors{2}{gray!25}{white}

        \begin{tabular}{lttt}
            \toprule
            & & \multicolumn{2}{c}{Jeffries prior} \\
            \cmidrule{3-4}
            & \multicolumn{1}{c}{$\AIC$ difference} & \multicolumn{1}{c}{$|\ln K|$, cutoff=1} & \multicolumn{1}{c}{$|\ln K|$, cutoff=0.01} \\
            $M_1$ vs. $M_3$ & -0.94 & 1.26 &0.75\\
            $M_2$ vs. $M_3$ & 0.62 & 0.64 &0.38\\
            \bottomrule
        \end{tabular}
    \end{center}
    The Akaike information criterion prefers $M_1$ to $M_3$, but $M_3$ to $M_2$.
    Therefore the trivial Gaussian model $M_1$ is
    preferred overall, but if we discard this
    option then the information criterion
    prefers a two-parameter fit ($M_3$) to a single-parameter fit ($M_2$).
    
    The Bayes factors are inconclusive, but it could be argued that
    they show a weak preference for the opposite conclusion.
    We compute the Bayes factor using two
    different choices for the regularization of the
    Jeffries prior; see footnote~\ref{footnote:jeffries}
    on p.~\pageref{footnote:jeffries}.
    A \emph{smaller} cutoff increases the weight of
    probability for the $\lambda_\alpha$ to be
    near zero, and therefore \emph{decreases} the
    probability that the model generates an observable
    signature.
    As we increase the lower limit for the parameters
    $\lambda_\alpha$ to the `natural' level
    $\lambda_\alpha=1$, 
    the Bayes factor does not strongly discriminate between
    a two- or three-parameter fit.
    However, it does
    marginally begin to disfavour
    a two-parameter fit ($M_3$) compared to the
    trivial model ($M_1$).
    Therefore it appears that a fit
    for a DBI-like model using
    more than
    two parameters becomes mildly in tension
    with the data for `natural' choices of the
    dimensionless scales $\lambda_\alpha$.

    The apparent discrepancy with the Akaike information criterion
    should be ascribed to a stronger `Ockham' or complexity penalty
    in the Bayes factor. The information criterion down-weights
    each model by a fixed amount depending on the number of parameters,
    whereas the Bayes factor attempts to account for the
    increased volume of parameter space which becomes available.
    For example, using a flat prior instead of the Jeffries
    prior very strongly disfavours the models $M_2$ and $M_3$.    
\end{itemize}

\section{Discussion and conclusions}
\label{sec:conclusions}

The availability of high-quality maps of the
CMB temperature anisotropy from the WMAP and Planck
missions means that it has become feasible to
search for primordial three-point correlations.
Such correlations are typically
predicted by any scenario in which the
fluctuations have an inflationary
origin, due to microphysical three-body interactions
among the light, active degrees of freedom
of the inflationary epoch.
If detected, their precise form could provide decisive
evidence in favour of the inflationary hypothesis.

Unfortunately, due to issues of computational complexity,
it is not yet possible to perform a blind search
for these primordial three-point correlations.
Instead, we must search for signals which we have
some prior reason to believe may be present in
the data.
Therefore
the amount of information we manage to extract depends
on which signals we choose to look for.

In this paper we have made a systematic search
of the 9-year WMAP data
for correlations which could be produced in a very
general model of single-field inflation,
under the assumption that the background evolution is
smooth, yielding corresponding smooth
and nearly scale-invariant correlation
functions. This excludes models which contain sharp
features or oscillations~\cite{Starobinsky:1992ts,
Adams:1997de,Adams:2001vc,Hailu:2006uj,
Bean:2008na,Achucarro:2010da,Joy:2007na,
Hotchkiss:2009pj,Nakashima:2010sa,Adshead:2011bw}.
It also excludes models in which significant three-point
correlations
are generated by differences of evolution
between regions of the universe separated by
super-Hubble distances.
Correlations generated by this mechanism
are generally most significant in the `squeezed'
or soft limit, where the correlation is between
fluctuations on very disparate scales.
Such correlations have been disfavoured
by analysis of the Planck2013 data release~\cite{Ade:2013ydc}.
By comparison, the 9-year WMAP data achieve a smaller
signal-to-noise for such configurations. The difference between the 9-year WMAP and Planck2013
datasets is less pronounced for the momentum configurations
which we probe, with for example $1\sigma$ error bars on $\fNL^{\rm{equil}}$ improving from $117$ to $75$.

The essential steps of our analysis
were assembled in~{\S\S\ref{sec:EFT}--\ref{sec:estimating}}.
We begin with an effective field theory which parametrizes
the unknown details of three-body interactions between
inflaton fluctuations, but preserves nonlinearly
realized Lorentz invariance.
The effective theory is agnostic regarding
the physical mechanism which underlies inflation.
We compute the bispectrum generated by
each operator in the effective theory,
and break these into principal components using
a Fisher-matrix approach.
The amplitude of each principal component is
recovered from the data, after which the results
can be translated into constraints on the
mass scales which appeared in the original
effective theory.
We find that no significant deviation from
Gaussianity has been detected in any region of the
inflationary parameter space.
This conclusion is consistent with previous
analyses of the 9-year WMAP and Planck2013 datasets.

Our principal components are similar to those
obtained by Byun \& Bean, who forecast the constraints
which could be obtained from a Planck-like survey~\cite{Byun:2013jba}.
We find that the best-constrained principal direction
exhibits similarities to
(in order) the flattened, orthogonal and `Galileon' templates.
A fourth principal direction is more complex, but at best
weakly constrained.

The large space of models
which fit into the class of single-field scenarios
invites attempts to
identify best-fitting regions.
To approach this problem we use the framework
of Bayesian model comparison.
The results are at best weakly significant,
but tend to disfavour models
with more parameters when compared to simpler
cases with zero or one parameter.
This is not surprising
given that the amplitude of each principal
direction is consistent with zero.
However, it should be borne in mind that our analysis
is restricted to smooth and nearly scale-invariant
bispectra.
It is possible that a significant signal
of a different type
is hidden in
the data.
In some cases,
$n$-point functions of this type
can be described within
the framework of effective field theory~\cite{Bartolo:2013exa}.
The analysis developed in~{\S\S\ref{sec:EFT}--\ref{sec:estimating}}
could be applied immediately to such scenarios
given a suitable choice of basis functions $\Rmode_n$.

\section*{Acknowledgements}
It is a pleasure to thank
Andrew Liddle for many helpful discussions and useful comments on an advance draft of this manuscript.
We are also grateful to Raquel Ribeiro and S\'{e}bastien Renaux-Petel for discussions which helped motivate
the approach we developed in this work. 

Some numerical presented in this paper were obtained using
the COSMOS supercomputer,
which is funded by STFC, HEFCE and SGI.
Other
numerical computations were carried out on the Sciama High Performance Compute (HPC)
cluster which is supported by
the ICG, SEPNet and the University of Portsmouth.
DR and DS acknowledge support from the Science and Technology
Facilities Council [grant number ST/I000976/1]. GA is supported by STFC grant ST/I506029/1.
DS also acknowledges support from the Leverhulme Trust.
The research leading to these results has received funding from
the European Research Council under the European Union's
Seventh Framework Programme (FP/2007--2013) / ERC Grant
Agreement No. [308082].

\appendix

\section{Three-point functions for the EFT operators}
\label{sec:appendbispec}
In this Appendix we briefly recapitulate the calculation
of the three-point functions corresponding
to each operator in the effective Lagrangian~\eqref{EFTLag}.
The principal tool is Schwinger's formulation
of `in--in' expectation values and the corresponding
expansion into diagrams due to
Keldysh~\cite{Schwinger:1960qe, Bakshi:1962dv,Bakshi:1963bn,Keldysh:1964ud}.
The technique was applied to general relativity by
Jordan, who used it to study the
effective equations of motion
obtained by integrating out quantum fluctuations~\cite{Jordan:1986ug}.
It was imported into cosmology by Calzetta \& Hu~\cite{Calzetta:1986ey}
and applied to inflation by Maldacena and subsequent
authors~\cite{Maldacena:2002vr,Weinberg:2005vy,Weinberg:2006ac}.

\para{In--in calculations}%
The objective is to calculate the expectation value
of a given operator $\Op$
at some time $t_\ast$,
given that the system develops from a specified
state (the `in'-state) at very early times.
This expectation value is
\begin{equation}
    \langle \Op \rangle_\ast \equiv
    \langle \text{in} | \Op(t_\ast) | \text{in} \rangle ,
\end{equation}
where the subscript `$\ast$' is used to denote evaluation
of $\Op$ at time $t_\ast$.
In the present case, $\Op$ will correspond to a product
of field operators evaluated at the same time but
at distinct spatial positions.

Inserting a complete set of intermediate states
labelled by the three-dimensional field configuration
$\pi(\vect{x}, T)$
at some arbitrary time $T > t_\ast$, we conclude
\begin{equation}
    \langle \Op \rangle_\ast
    =
    \int [ \d \pi(\vect{x},T) ]
    \,
    \langle \text{in} | \pi(\vect{x},T) \rangle
    \langle \pi(\vect{x},T) | \Op(t_\ast) | \text{in} \rangle ,
    \label{eq:keldysh-primitive}
\end{equation}
where the measure $[\d \pi(\vect{x},T) ]$ denotes
integration over all field configurations.
Each overlap in~\eqref{eq:keldysh-primitive}
can be written as a conventional Feynman path integral,
with the integration running over all field
\emph{histories}
$\pi(\vect{x},t)$ which are consistent
with the in-state $| \text{in} \rangle$ in the
far past, and which coincide with the
configuration
$\pi(\vect{x},T)$ at time $T$.
The result is
\begin{equation}
    \langle \Op(t_{\ast}) \rangle
    = \int [\d\pi_{+}\,\d\pi_{-}]
        \,
        \Op(t_\ast)
        \,
        \exp\Big[
            \im S(\pi_{+})
            - \im S(\pi_{-})
        \Big] \,
        \delta[\pi_{+}(T) - \pi_{-}(T)] ,
    \label{eq:keldysh-path-integral}
\end{equation}
with the independent integrations
$\pi_+$, $\pi_-$ running over field
histories which are compatible with the in-state
but are unrestricted at late times.
Eq.~\eqref{eq:keldysh-path-integral}
admits an expansion into diagrams
in which the Green's functions
connecting only `$+$' or only `$-$' fields
obey the usual Feynman boundary conditions,
but are augmented by Green's
functions which mix the `$+$' and `$-$'
labels and whose boundary conditions are
determined by the $\delta$-function.
For further details, see Ref.~\cite{Weinberg:2005vy}.

\para{Mode functions}%
It was explained in~{\S\ref{sec:bispec}}
that we approximate the mode functions
as Hankel functions of order 3/2.
Analytically, this corresponds to building
Green's functions from the mode function
\begin{equation}
    \label{modefunc}
    u(\tau,\vect{k})= \frac{\im H}{\sqrt{4 \epsilon \tildecs k^3}}(1+\im k \tildecs \tau)e^{-\im k \tildecs \tau},
\end{equation}
and its complex conjugate.
In this formula, $\tau = -\int_t^{\infty} \d t'/a(t')$
is the conformal time
and $\tildecs$ is a `generalized' speed of sound.
In a model without fourth-derivative
kinetic terms this will usually be the phase velocity,
determined from the ratio of coefficients
of the spatial and temporal kinetic terms.
In other cases it may bear less relation to what would
normally be thought of as a phase velocity.
Our notation coincides with that of
Refs.~\cite{Bartolo:2010bj,Bartolo:2010di},
to which we refer for further details;
see especially the discussion below
Eq. (2.6) in Ref.~\cite{Bartolo:2010di}.
In writing Eq.~\eqref{modefunc}
we have assumed that the in-state
$| \text{in} \rangle$ contains zero particles,
corresponding the `Bunch--Davies' vacuum.

With these choices, the
three-point functions corresponding to the
EFT operators in~\eqref{EFTLag}
are:
\begin{itemize} 
\item $\mathcal{O}_{A}= - \bar{M}_{1}^3 (\partial \pi)^2 \partial^{2}\pi/2a^4$
\[  B_{\zeta}(k_1,k_2,k_3) \supseteq \frac{1}{16}\bar{M}_{1}^{3}\frac{H_{\star}^{3}}{\epsilon^{3}{c}_{s}^{4}\prod\limits_{i}k_{i}^{3}}\,k_{1}^{2} \,\, \vect{k}_{2}.\vect{k}_{3}\,\,  \mathcal{A}_{1} + 1 \!\rightarrow \! 2 + 1 \! \rightarrow \! 3\,,\hspace{3.5cm} \]
\item $\mathcal{O}_{B} = -2 M_{2}^4\dot{\pi} (\partial \pi)^2/a^2$
 \begin{multline*}  B_{\zeta}(k_1,k_2,k_3)  \supseteq  \frac{1}{8} M_{2}^{4}\frac{H_{\star}^{2}}{\epsilon^{3}{c}_{s}^{2}\prod\limits_{i}k_{i}^{3}} k_{1}^{2}\, \vect{k}_{2}.\vect{k}_{3}\left(\frac{1}{k_{t}} + \frac{k_{2}+k_{3}}{k_{t}^2} + \frac{2k_{2}k_{3}}{k_{t}^3}\right)   + 1 \! \rightarrow \! 2 + 1\! \rightarrow \! 3\,, \end{multline*}
\item $\mathcal{O}_{C} = - \bar{M}_{2}^2 \left[ H(\partial^{2}\pi)(\partial \pi)^2/2 +  \dot{\pi}\partial^{2}\partial_{j}\pi\partial_{j}\pi \right]/a^4$
 \begin{multline*} B_{\zeta}(k_1,k_2,k_3)\supseteq \frac{1}{16}  \bar{M}_{2}^{2}\frac{H_{\star}^{4}}{\epsilon^{3}{c}_{s}^{4}\prod\limits_{i}k_{i}^{3}}\,  k_{1}^{2} \,\, \vect{k}_{2}.\vect{k}_{3}  \left[ \mathcal{A}_{1} + (k_{2}^{2}+k_{3}^{2}) \,\,\mathcal{A}_{2}\right] + 1\!  \rightarrow \! 2 + 1\! \rightarrow \! 3\,, \end{multline*}
\item $\mathcal{O}_{D} = - 4 M_{3}^4 \dot{\pi}^3/3$
\[B_{\zeta}(k_1,k_2,k_3)\supseteq \frac{1}{2}  M_{3}^{4}\frac{H_{\star}^{2}}{\epsilon^{3}\prod\limits_{i}k_{i}} \frac{1}{k_{t}^3}\,,\hspace{8cm} \]
\item $\mathcal{O}_{E}= - \bar{M}_{3}^{2} \left[ H (\partial \pi)^{2}\partial^{2}\pi + \dot{\pi}\partial^{2}\partial_{j}\pi\partial_{j} \pi \right] /a^4$
\begin{multline*} B_{\zeta}(k_1,k_2,k_3)\supseteq \frac{1}{8}  \bar{M}_{3}^{2}\frac{H_{\star}^{4}}{\epsilon^{3} {c}_{s}^{4} \prod\limits_{i}k_{i}^{3}} \, k_{1}^{2}\,\,\vect{k}_{2}.\vect{k}_{3} \,\, \left(\mathcal{A}_{1}+\frac{k_2^2+k_3^2}{2}\mathcal{A}_{2}\right) + 1\!  \rightarrow \! 2 + 1\! \rightarrow \! 3\,,\end{multline*}
 \item $\mathcal{O}_{F}= -2 \bar{M}_{4}^3\dot{\pi}^{2}\partial^{2}\pi/ 3a^2$ 
 \[ B_{\zeta}(k_1,k_2,k_3)\supseteq \frac{1}{2}   \bar{M}_{4}^{3}\frac{H_{\star}^{3}}{\epsilon^{3}{c}_{s}^{2}\prod\limits_{i}k_{i}}\frac{1}{k_{t}^{3}}\,,\hspace{7.5cm} \]
  \item $\mathcal{O}_{G} = \bar{M}_{5}^{2}\dot{\pi}(\partial^{2}\pi)^{2}/3a^4$
  \[ B_{\zeta}(k_1,k_2,k_3)\supseteq -\frac{1}{8}  \bar{M}_{5}^{2}\frac{H_{\star}^{4}}{\epsilon^{3} {c}_{s}^{4} \prod\limits_{i}k_{i}}\, \frac{1}{k_{t}^{3}}\left(3 + \frac{4k^{2}_{\alpha}}{k_{t}^2} \right)\,, \hspace{5.5cm}\]
  \item $\mathcal{O}_{H}=\bar{M}_{6}^{2}\dot{\pi}(\partial_{i}\partial_{j}\pi)^{2}/3a^4 $ 
\[ B_{\zeta}(k_1,k_2,k_3)\supseteq -\frac{1}{24}  \bar{M}_{6}^{2}\frac{H_{\star}^{4}}{\epsilon^{3} {c}_{s}^{4}\prod\limits_{i}k_{i}^{3}}\,  k_{1}^{2} \,\, (\vect{k}_{2}.\vect{k}_{3})^{2} \,\, \mathcal{A}_{2}  \,\, + 1\!  \rightarrow \! 2 + 1\! \rightarrow \! 3\,, \hspace{3.5cm}\]
\item $\mathcal{O}_{I}=- \bar{M}_{7} (\partial^{2}\pi)^{3}/3!a^6$
  \[ B_{\zeta}(k_1,k_2,k_3)\supseteq \frac{1}{4}   \bar{M}_{7} \frac{H_{\star}^{5}}{\epsilon^{3} {c}_{s}^{6} \prod\limits_{i}k_{i}}\,  \, \mathcal{A}_{3}\,, \hspace{7.5cm}\]
\item $\mathcal{O}_{J}= - \bar{M}_{8}\partial^{2}\pi (\partial_{j}\partial_{k}\pi)^2/ 3! a^6$  
  \[ B_{\zeta}(k_1,k_2,k_3)\supseteq \frac{1}{12}  \bar{M}_{8} \frac{H_{\star}^{5}}{\epsilon^{3} {c}_{s}^{6}\prod\limits_{i}k_{i}^{3}}\,   k_{1}^{2}\,\, (\vect{k}_{2}. \vect{k}_{3})^{2} \,\, \mathcal{A}_{3}+ 1\!  \rightarrow \! 2 + 1\! \rightarrow \! 3\,, \hspace{2.5cm}\]
  \item $\mathcal{O}_{K}= - \bar{M}_{9}\partial_{i}\partial_{j}\pi \partial_{j}\partial_{k}\pi \partial_{k}\partial_{i}\pi/ 3! a^6$
 \[ B_{\zeta}(k_1,k_2,k_3)\supseteq \frac{1}{4}   \bar{M}_{9} \frac{H_{\star}^{5}}{\epsilon^{3} {c}_{s}^{6} \prod\limits_{i}k_{i}^{3}}\,  \, (\vect{k}_{1}. \vect{k}_{2})(\vect{k}_{1}. \vect{k}_{3})(\vect{k}_{2}. \vect{k}_{3})  \,\, \mathcal{A}_{3}\,, \hspace{3.5cm}\]
\end{itemize}
where 
\begin{eqnarray*}
\mathcal{A}_{1} =\left( \frac{1}{k_{t}} + \frac{k^{2}_{\alpha}}{k_{t}^3} + \frac{3k^{3}_{\beta}}{k_{t}^4} \right), \quad
\mathcal{A}_{2} = \left(\frac{1}{k_{t}^{3}} + \frac{3(k_{2}+k_{3})}{k_{t}^4} + \frac{12k_{2}k_{3}}{k_{t}^{5}}\right), \quad
\mathcal{A}_{3} = \left(\frac{1}{k_{t}^{3}} + \frac{3k^{2}_{\alpha}}{k_{t}^5} + \frac{15k^{3}_{\beta}}{k_{t}^{6}}\right),
\end{eqnarray*}
with $k_t = k_1 + k_2 + k_3$, $\,\,k^{2}_{\alpha} = k_1k_2 + k_1k_3 + k_2k_3$, and $k^{3}_\beta = k_1k_2k_3$.

\bibliography{bibli}

\providecommand{\href}[2]{#2}\begingroup\raggedright\begin{thebibliography}{10}

\bibitem{Ade:2013ydc}
{\bf Planck Collaboration} Collaboration, P.~Ade {\em et~al.}, {\it {Planck
  2013 Results. XXIV. Constraints on primordial non-Gaussianity}},
  \href{http://arxiv.org/abs/1303.5084}{{\tt arXiv:1303.5084}}.

\bibitem{Hinshaw:2012aka}
{\bf WMAP Collaboration} Collaboration, G.~Hinshaw {\em et~al.}, {\it
  {Nine-Year Wilkinson Microwave Anisotropy Probe (WMAP) Observations:
  Cosmological Parameter Results}},  \href{http://arxiv.org/abs/1212.5226}{{\tt
  arXiv:1212.5226}}.

\bibitem{Cheung:2007st}
C.~Cheung, P.~Creminelli, A.~L. Fitzpatrick, J.~Kaplan, and L.~Senatore, {\it
  {The Effective Field Theory of Inflation}},  {\em JHEP} {\bf 0803} (2008)
  014, [\href{http://arxiv.org/abs/0709.0293}{{\tt arXiv:0709.0293}}].

\bibitem{Weinberg:2003sw}
S.~Weinberg, {\it {Adiabatic modes in cosmology}},  {\em Phys.Rev.} {\bf D67}
  (2003) 123504, [\href{http://arxiv.org/abs/astro-ph/0302326}{{\tt
  astro-ph/0302326}}].

\bibitem{Weinberg:2004kr}
S.~Weinberg, {\it {Can non-adiabatic perturbations arise after single-field
  inflation?}},  {\em Phys.Rev.} {\bf D70} (2004) 043541,
  [\href{http://arxiv.org/abs/astro-ph/0401313}{{\tt astro-ph/0401313}}].

\bibitem{Weinberg:2004kf}
S.~Weinberg, {\it {Must cosmological perturbations remain non-adiabatic after
  multi-field inflation?}},  {\em Phys.Rev.} {\bf D70} (2004) 083522,
  [\href{http://arxiv.org/abs/astro-ph/0405397}{{\tt astro-ph/0405397}}].

\bibitem{Meyers:2010rg}
J.~Meyers and N.~Sivanandam, {\it {Non-Gaussianities in Multifield Inflation:
  Superhorizon Evolution, Adiabaticity, and the Fate of fnl}},  {\em Phys.Rev.}
  {\bf D83} (2011) 103517, [\href{http://arxiv.org/abs/1011.4934}{{\tt
  arXiv:1011.4934}}].

\bibitem{Elliston:2011dr}
J.~Elliston, D.~J. Mulryne, D.~Seery, and R.~Tavakol, {\it {Evolution of fNL to
  the adiabatic limit}},  {\em JCAP} {\bf 1111} (2011) 005,
  [\href{http://arxiv.org/abs/1106.2153}{{\tt arXiv:1106.2153}}].

\bibitem{Creminelli:2006xe}
P.~Creminelli, M.~A. Luty, A.~Nicolis, and L.~Senatore, {\it {Starting the
  Universe: Stable Violation of the Null Energy Condition and Non-standard
  Cosmologies}},  {\em JHEP} {\bf 0612} (2006) 080,
  [\href{http://arxiv.org/abs/hep-th/0606090}{{\tt hep-th/0606090}}].

\bibitem{Cheung:2007sv}
C.~Cheung, A.~L. Fitzpatrick, J.~Kaplan, and L.~Senatore, {\it {On the
  consistency relation of the 3-point function in single field inflation}},
  {\em JCAP} {\bf 0802} (2008) 021, [\href{http://arxiv.org/abs/0709.0295}{{\tt
  arXiv:0709.0295}}].

\bibitem{Weinberg:2008hq}
S.~Weinberg, {\it {Effective Field Theory for Inflation}},  {\em Phys.Rev.}
  {\bf D77} (2008) 123541, [\href{http://arxiv.org/abs/0804.4291}{{\tt
  arXiv:0804.4291}}].

\bibitem{Senatore:2009gt}
L.~Senatore, K.~M. Smith, and M.~Zaldarriaga, {\it {Non-Gaussianities in Single
  Field Inflation and their Optimal Limits from the WMAP 5-year Data}},  {\em
  JCAP} {\bf 1001} (2010) 028, [\href{http://arxiv.org/abs/0905.3746}{{\tt
  arXiv:0905.3746}}].

\bibitem{Bartolo:2010bj}
N.~Bartolo, M.~Fasiello, S.~Matarrese, and A.~Riotto, {\it {Large
  non-Gaussianities in the Effective Field Theory Approach to Single-Field
  Inflation: the Bispectrum}},  {\em JCAP} {\bf 1008} (2010) 008,
  [\href{http://arxiv.org/abs/1004.0893}{{\tt arXiv:1004.0893}}].

\bibitem{Bartolo:2010di}
N.~Bartolo, M.~Fasiello, S.~Matarrese, and A.~Riotto, {\it {Large
  non-Gaussianities in the Effective Field Theory Approach to Single-Field
  Inflation: the Trispectrum}},  {\em JCAP} {\bf 1009} (2010) 035,
  [\href{http://arxiv.org/abs/1006.5411}{{\tt arXiv:1006.5411}}].

\bibitem{Bartolo:2013exa}
N.~Bartolo, D.~Cannone, and S.~Matarrese, {\it {The Effective Field Theory of
  Inflation Models with Sharp Features}},  {\em JCAP} {\bf 1310} (2013) 038,
  [\href{http://arxiv.org/abs/1307.3483}{{\tt arXiv:1307.3483}}].

\bibitem{Adshead:2013zfa}
P.~Adshead, W.~Hu, and V.~Miranda, {\it {Bispectrum in Single-Field Inflation
  Beyond Slow-Roll}},  {\em Phys.Rev.} {\bf D88} (2013) 023507,
  [\href{http://arxiv.org/abs/1303.7004}{{\tt arXiv:1303.7004}}].

\bibitem{Alishahiha:2004eh}
M.~Alishahiha, E.~Silverstein, and D.~Tong, {\it {DBI in the sky}},  {\em
  Phys.Rev.} {\bf D70} (2004) 123505,
  [\href{http://arxiv.org/abs/hep-th/0404084}{{\tt hep-th/0404084}}].

\bibitem{ArkaniHamed:2003uz}
N.~Arkani-Hamed, P.~Creminelli, S.~Mukohyama, and M.~Zaldarriaga, {\it {Ghost
  inflation}},  {\em JCAP} {\bf 0404} (2004) 001,
  [\href{http://arxiv.org/abs/hep-th/0312100}{{\tt hep-th/0312100}}].

\bibitem{Dias:2012qy}
M.~Dias, R.~H. Ribeiro, and D.~Seery, {\it {The δN formula is the dynamical
  renormalization group}},  {\em JCAP} {\bf 1310} (2013) 062,
  [\href{http://arxiv.org/abs/1210.7800}{{\tt arXiv:1210.7800}}].

\bibitem{Assassi:2012et}
V.~Assassi, D.~Baumann, and D.~Green, {\it {Symmetries and Loops in
  Inflation}},  {\em JHEP} {\bf 1302} (2013) 151,
  [\href{http://arxiv.org/abs/1210.7792}{{\tt arXiv:1210.7792}}].

\bibitem{Senatore:2009cf}
L.~Senatore and M.~Zaldarriaga, {\it {On Loops in Inflation}},  {\em JHEP} {\bf
  1012} (2010) 008, [\href{http://arxiv.org/abs/0912.2734}{{\tt
  arXiv:0912.2734}}].

\bibitem{Senatore:2010jy}
L.~Senatore and M.~Zaldarriaga, {\it {A Naturally Large Four-Point Function in
  Single Field Inflation}},  {\em JCAP} {\bf 1101} (2011) 003,
  [\href{http://arxiv.org/abs/1004.1201}{{\tt arXiv:1004.1201}}].

\bibitem{Senatore:2010wk}
L.~Senatore and M.~Zaldarriaga, {\it {The Effective Field Theory of Multifield
  Inflation}},  {\em JHEP} {\bf 1204} (2012) 024,
  [\href{http://arxiv.org/abs/1009.2093}{{\tt arXiv:1009.2093}}].

\bibitem{Baumann:2011su}
D.~Baumann and D.~Green, {\it {Equilateral Non-Gaussianity and New Physics on
  the Horizon}},  {\em JCAP} {\bf 1109} (2011) 014,
  [\href{http://arxiv.org/abs/1102.5343}{{\tt arXiv:1102.5343}}].

\bibitem{Baumann:2011dt}
D.~Baumann, L.~Senatore, and M.~Zaldarriaga, {\it {Scale-Invariance and the
  Strong Coupling Problem}},  {\em JCAP} {\bf 1105} (2011) 004,
  [\href{http://arxiv.org/abs/1101.3320}{{\tt arXiv:1101.3320}}].

\bibitem{Maldacena:2002vr}
J.~M. Maldacena, {\it {Non-Gaussian features of primordial fluctuations in
  single field inflationary models}},  {\em JHEP} {\bf 05} (2003) 013,
  [\href{http://arxiv.org/abs/astro-ph/0210603}{{\tt astro-ph/0210603}}].

\bibitem{Creminelli:2003iq}
P.~Creminelli, {\it {On non-Gaussianities in single-field inflation}},  {\em
  JCAP} {\bf 0310} (2003) 003,
  [\href{http://arxiv.org/abs/astro-ph/0306122}{{\tt astro-ph/0306122}}].

\bibitem{Seery:2005wm}
D.~Seery and J.~E. Lidsey, {\it {Primordial non-gaussianities in single field
  inflation}},  {\em JCAP} {\bf 0506} (2005) 003,
  [\href{http://arxiv.org/abs/astro-ph/0503692}{{\tt astro-ph/0503692}}].

\bibitem{Seery:2005gb}
D.~Seery and J.~E. Lidsey, {\it {Primordial non-gaussianities from
  multiple-field inflation}},  {\em JCAP} {\bf 0509} (2005) 011,
  [\href{http://arxiv.org/abs/astro-ph/0506056}{{\tt astro-ph/0506056}}].

\bibitem{Weinberg:2005vy}
S.~Weinberg, {\it {Quantum contributions to cosmological correlations}},  {\em
  Phys.Rev.} {\bf D72} (2005) 043514,
  [\href{http://arxiv.org/abs/hep-th/0506236}{{\tt hep-th/0506236}}].

\bibitem{Chen:2006nt}
X.~Chen, M.-x. Huang, S.~Kachru, and G.~Shiu, {\it {Observational signatures
  and non-Gaussianities of general single field inflation}},  {\em JCAP} {\bf
  0701} (2007) 002, [\href{http://arxiv.org/abs/hep-th/0605045}{{\tt
  hep-th/0605045}}].

\bibitem{Burrage:2011hd}
C.~Burrage, R.~H. Ribeiro, and D.~Seery, {\it {Large slow-roll corrections to
  the bispectrum of noncanonical inflation}},  {\em JCAP} {\bf 1107} (2011)
  032, [\href{http://arxiv.org/abs/1103.4126}{{\tt arXiv:1103.4126}}].

\bibitem{Elliston:2012ab}
J.~Elliston, D.~Seery, and R.~Tavakol, {\it {The inflationary bispectrum with
  curved field-space}},  {\em JCAP} {\bf 1211} (2012) 060,
  [\href{http://arxiv.org/abs/1208.6011}{{\tt arXiv:1208.6011}}].

\bibitem{Fergusson:2009nv}
J.~Fergusson, M.~Liguori, and E.~Shellard, {\it {General CMB and Primordial
  Bispectrum Estimation I: Mode Expansion, Map-Making and Measures of
  $f_{NL}$}},  {\em Phys.Rev.} {\bf D82} (2010) 023502,
  [\href{http://arxiv.org/abs/0912.5516}{{\tt arXiv:0912.5516}}].

\bibitem{Fergusson:2010dm}
J.~Fergusson, M.~Liguori, and E.~Shellard, {\it {The CMB Bispectrum}},  {\em
  JCAP} {\bf 1212} (2012) 032, [\href{http://arxiv.org/abs/1006.1642}{{\tt
  arXiv:1006.1642}}].

\bibitem{Fergusson:2010ia}
J.~Fergusson, D.~Regan, and E.~Shellard, {\it {Rapid Separable Analysis of
  Higher Order Correlators in Large Scale Structure}},  {\em Phys.Rev.} {\bf
  D86} (2012) 063511, [\href{http://arxiv.org/abs/1008.1730}{{\tt
  arXiv:1008.1730}}].

\bibitem{Fergusson:2010gn}
J.~Fergusson, D.~Regan, and E.~Shellard, {\it {Optimal Trispectrum Estimators
  and WMAP Constraints}},  \href{http://arxiv.org/abs/1012.6039}{{\tt
  arXiv:1012.6039}}.

\bibitem{Regan:2010cn}
D.~Regan, E.~Shellard, and J.~Fergusson, {\it {General CMB and Primordial
  Trispectrum Estimation}},  {\em Phys.Rev.} {\bf D82} (2010) 023520,
  [\href{http://arxiv.org/abs/1004.2915}{{\tt arXiv:1004.2915}}].

\bibitem{Byun:2013jba}
J.~Byun and R.~Bean, {\it {Non-Gaussian Shape Recognition}},  {\em JCAP} {\bf
  1309} (2013) 026, [\href{http://arxiv.org/abs/1303.3050}{{\tt
  arXiv:1303.3050}}].

\bibitem{Battefeld:2011ut}
T.~Battefeld and J.~Grieb, {\it {Anatomy of bispectra in general single-field
  inflation -- modal expansions}},  {\em JCAP} {\bf 1112} (2011) 003,
  [\href{http://arxiv.org/abs/1110.1369}{{\tt arXiv:1110.1369}}].

\bibitem{Regan:2013wwa}
D.~Regan, P.~Mukherjee, and D.~Seery, {\it {General CMB bispectrum analysis
  using wavelets and separable modes}},
  \href{http://arxiv.org/abs/1302.5631}{{\tt arXiv:1302.5631}}.

\bibitem{Regan:2013jua}
D.~Regan, M.~Gosenca, and D.~Seery, {\it {Constraining the WMAP9 bispectrum and
  trispectrum with needlets}},  \href{http://arxiv.org/abs/1310.8617}{{\tt
  arXiv:1310.8617}}.

\bibitem{Fergusson:2008ra}
J.~Fergusson and E.~Shellard, {\it {The shape of primordial non-Gaussianity and
  the CMB bispectrum}},  {\em Phys.Rev.} {\bf D80} (2009) 043510,
  [\href{http://arxiv.org/abs/0812.3413}{{\tt arXiv:0812.3413}}].

\bibitem{Vallisneri:2007ev}
M.~Vallisneri, {\it {Use and abuse of the Fisher information matrix in the
  assessment of gravitational-wave parameter-estimation prospects}},  {\em
  Phys.Rev.} {\bf D77} (2008) 042001,
  [\href{http://arxiv.org/abs/gr-qc/0703086}{{\tt gr-qc/0703086}}].

\bibitem{Babich:2004gb}
D.~Babich, P.~Creminelli, and M.~Zaldarriaga, {\it {The Shape of
  non-Gaussianities}},  {\em JCAP} {\bf 0408} (2004) 009,
  [\href{http://arxiv.org/abs/astro-ph/0405356}{{\tt astro-ph/0405356}}].

\bibitem{Meerburg:2009ys}
P.~D. Meerburg, J.~P. van~der Schaar, and P.~S. Corasaniti, {\it {Signatures of
  Initial State Modifications on Bispectrum Statistics}},  {\em JCAP} {\bf
  0905} (2009) 018, [\href{http://arxiv.org/abs/0901.4044}{{\tt
  arXiv:0901.4044}}].

\bibitem{Creminelli:2010qf}
P.~Creminelli, G.~D'Amico, M.~Musso, J.~Norena, and E.~Trincherini, {\it
  {Galilean symmetry in the effective theory of inflation: new shapes of
  non-Gaussianity}},  {\em JCAP} {\bf 1102} (2011) 006,
  [\href{http://arxiv.org/abs/1011.3004}{{\tt arXiv:1011.3004}}].

\bibitem{Ribeiro:2011ax}
R.~H. Ribeiro and D.~Seery, {\it {Decoding the bispectrum of single-field
  inflation}},  {\em JCAP} {\bf 1110} (2011) 027,
  [\href{http://arxiv.org/abs/1108.3839}{{\tt arXiv:1108.3839}}].

\bibitem{Bennett:2012zja}
{\bf WMAP} Collaboration, C.~Bennett {\em et~al.}, {\it {Nine-Year Wilkinson
  Microwave Anisotropy Probe (WMAP) Observations: Final Maps and Results}},
  {\em Astrophys.J.Suppl.} {\bf 208} (2013) 20,
  [\href{http://arxiv.org/abs/1212.5225}{{\tt arXiv:1212.5225}}].

\bibitem{Martin:2013nzq}
J.~Martin, C.~Ringeval, R.~Trotta, and V.~Vennin, {\it {The Best Inflationary
  Models After Planck}},  \href{http://arxiv.org/abs/1312.3529}{{\tt
  arXiv:1312.3529}}.

\bibitem{doi:10.1080/01621459.1995.10476572}
R.~E. Kass and A.~E. Raftery, {\it Bayes factors},  {\em Journal of the
  American Statistical Association} {\bf 90} (1995), no.~430 773--795,
  [\href{http://arxiv.org/abs/http://amstat.tandfonline.com/doi/pdf/10.1080/01621459.1995.10476572}{{\tt
  http://amstat.tandfonline.com/doi/pdf/10.1080/01621459.1995.10476572}}].

\bibitem{Starobinsky:1992ts}
A.~A. Starobinsky, {\it {Spectrum of adiabatic perturbations in the universe
  when there are singularities in the inflation potential}},  {\em JETP Lett.}
  {\bf 55} (1992) 489--494.

\bibitem{Adams:1997de}
J.~A. Adams, G.~G. Ross, and S.~Sarkar, {\it {Multiple inflation}},  {\em
  Nucl.Phys.} {\bf B503} (1997) 405--425,
  [\href{http://arxiv.org/abs/hep-ph/9704286}{{\tt hep-ph/9704286}}].

\bibitem{Adams:2001vc}
J.~A. Adams, B.~Cresswell, and R.~Easther, {\it {Inflationary perturbations
  from a potential with a step}},  {\em Phys.Rev.} {\bf D64} (2001) 123514,
  [\href{http://arxiv.org/abs/astro-ph/0102236}{{\tt astro-ph/0102236}}].

\bibitem{Hailu:2006uj}
G.~Hailu and S.-H.~H. Tye, {\it {Structures in the Gauge/Gravity Duality
  Cascade}},  {\em JHEP} {\bf 0708} (2007) 009,
  [\href{http://arxiv.org/abs/hep-th/0611353}{{\tt hep-th/0611353}}].

\bibitem{Bean:2008na}
R.~Bean, X.~Chen, G.~Hailu, S.-H.~H. Tye, and J.~Xu, {\it {Duality Cascade in
  Brane Inflation}},  {\em JCAP} {\bf 0803} (2008) 026,
  [\href{http://arxiv.org/abs/0802.0491}{{\tt arXiv:0802.0491}}].

\bibitem{Achucarro:2010da}
A.~Achucarro, J.-O. Gong, S.~Hardeman, G.~A. Palma, and S.~P. Patil, {\it
  {Features of heavy physics in the CMB power spectrum}},  {\em JCAP} {\bf
  1101} (2011) 030, [\href{http://arxiv.org/abs/1010.3693}{{\tt
  arXiv:1010.3693}}].

\bibitem{Joy:2007na}
M.~Joy, V.~Sahni, and A.~A. Starobinsky, {\it {A New Universal Local Feature in
  the Inflationary Perturbation Spectrum}},  {\em Phys.Rev.} {\bf D77} (2008)
  023514, [\href{http://arxiv.org/abs/0711.1585}{{\tt arXiv:0711.1585}}].

\bibitem{Hotchkiss:2009pj}
S.~Hotchkiss and S.~Sarkar, {\it {Non-Gaussianity from violation of slow-roll
  in multiple inflation}},  {\em JCAP} {\bf 1005} (2010) 024,
  [\href{http://arxiv.org/abs/0910.3373}{{\tt arXiv:0910.3373}}].

\bibitem{Nakashima:2010sa}
M.~Nakashima, R.~Saito, Y.-i. Takamizu, and J.~Yokoyama, {\it {The effect of
  varying sound velocity on primordial curvature perturbations}},  {\em
  Prog.Theor.Phys.} {\bf 125} (2011) 1035--1052,
  [\href{http://arxiv.org/abs/1009.4394}{{\tt arXiv:1009.4394}}].

\bibitem{Adshead:2011bw}
P.~Adshead, W.~Hu, C.~Dvorkin, and H.~V. Peiris, {\it {Fast Computation of
  Bispectrum Features with Generalized Slow Roll}},  {\em Phys.Rev.} {\bf D84}
  (2011) 043519, [\href{http://arxiv.org/abs/1102.3435}{{\tt
  arXiv:1102.3435}}].

\bibitem{Schwinger:1960qe}
J.~S. Schwinger, {\it {Brownian motion of a quantum oscillator}},  {\em
  J.Math.Phys.} {\bf 2} (1961) 407--432.

\bibitem{Bakshi:1962dv}
P.~M. Bakshi and K.~T. Mahanthappa, {\it {Expectation value formalism in
  quantum field theory. 1.}},  {\em J.Math.Phys.} {\bf 4} (1963) 1--11.

\bibitem{Bakshi:1963bn}
P.~M. Bakshi and K.~T. Mahanthappa, {\it {Expectation value formalism in
  quantum field theory. 2.}},  {\em J.Math.Phys.} {\bf 4} (1963) 12--16.

\bibitem{Keldysh:1964ud}
L.~Keldysh, {\it {Diagram technique for nonequilibrium processes}},  {\em
  Zh.Eksp.Teor.Fiz.} {\bf 47} (1964) 1515--1527.

\bibitem{Jordan:1986ug}
R.~Jordan, {\it {Effective Field Equations for Expectation Values}},  {\em
  Phys.Rev.} {\bf D33} (1986) 444--454.

\bibitem{Calzetta:1986ey}
E.~Calzetta and B.~Hu, {\it {Closed Time Path Functional Formalism in Curved
  Space-Time: Application to Cosmological Back Reaction Problems}},  {\em
  Phys.Rev.} {\bf D35} (1987) 495.

\bibitem{Weinberg:2006ac}
S.~Weinberg, {\it {Quantum contributions to cosmological correlations. II. Can
  these corrections become large?}},  {\em Phys.Rev.} {\bf D74} (2006) 023508,
  [\href{http://arxiv.org/abs/hep-th/0605244}{{\tt hep-th/0605244}}].

\end{thebibliography}\endgroup
\end{document}